\documentclass[apj]{emulateapj}

\usepackage{epsfig}
\usepackage{graphicx}
\usepackage{comment}
\usepackage{multirow}

\long\def\symbolfootnote[#1]#2{\begingroup%
\def\thefootnote{\fnsymbol{footnote}}\footnote[#1]{#2}\endgroup}

\def\figtxt{\footnotesize}
\def\myfont{\fontsize{10}{10} \selectfont}

\def\Om{$\Omega_M$}
\def\Olam{$\Omega_{\Lambda}$}
\def\um{$\mu$m}

\def\x{$\times$}
\def\twid{$\sim$}
\def\agnfrac{$f_{\mathrm{IR-AGN}}$}

\def\Mstar{$\mathrm{M}^*_{3.6}$}

\begin{document}

\title{A Census of Mid-Infrared Selected Active Galactic Nuclei in
  Massive Galaxy Clusters at $0 \lesssim \MakeLowercase{z} \lesssim
  1.3$}

\author{Adam R. Tomczak\footnotemark[1], Kim-Vy
H. Tran$^{1,2}$, and Am\'elie Saintonge\footnotemark[3]}
\footnotetext[1]{ George P. and Cynthia W. Mitchell Institute for
Fundamental Physics and Astronomy, Department of Physics and
Astronomy, Texas A\&M University, College Station, TX 77843 }
\footnotetext[2]{ Institute for Theoretical Physics, University of
Z\"urich, CH-8057, Switzerland }
\footnotetext[3]{ MPE: Max-Planck Institut f\"{u}r extraterrestrische
Physik, 85741 Garching, Germany } 

\begin{abstract} \myfont

We conduct a deep mid-infrared census of nine massive galaxy clusters
at $(0<z<1.3)$ with a total of $\sim1500$ spectroscopically confirmed
member galaxies using \textit{Spitzer}/IRAC photometry and established
mid-infrared color selection techniques.  Of the 949 cluster galaxies
that are detected in at least three of the four IRAC channels at the
$\geq 3\sigma$ level, we identify 12 that host mid-infrared selected
active galactic nuclei (IR-AGN).  To compare the IR-AGN across our
redshift range, we define two complete samples of cluster galaxies:
(1) optically-selected members with rest-frame $V_{\mathrm{AB}}$
magnitude $<-21.5$ and (2) mid-IR selected members brighter than
(\Mstar$+0.5)$, i.e. essentially a stellar mass cut.  In both samples,
we measure \agnfrac\ \twid\ 1\% with a strong upper limit of
\twid\ 3\% at $z<1$.
%, and $\lesssim5\%$ when considering only late-type members
This uniformly low IR-AGN fraction at $z<1$ is
surprising given the fraction of $24\mu$m sources in the same galaxy
clusters is observed to increase by about a factor of four from
$z\sim0$ to $z\sim1$; this indicates that most of the detected
$24\mu$m flux is due to star formation.  Only in our single galaxy
cluster at $z=1.24$ is the IR-AGN fraction measurably higher at
$\sim15$\% (all members; $\sim70$\% for late-types only).
%Furthermore, the \agnfrac\ is \twid 2\% for field galaxies in the
%AGES ($0.25 < z < 0.8$); this is larger than our cluster fractions at
%similar redshift, but still within the 1$\sigma$ uncertainties. 
In agreement with recent studies, we find the cluster IR-AGN are
predominantly hosted by late-type galaxies with blue optical colors, i.e. members
with recent/ongoing star formation.  The four brightest IR-AGN are
also X-ray sources; these IR+X-ray AGN all lie outside the cluster
core ($R_{proj}\gtrsim0.5$~Mpc) and are hosted by highly morphologically
disturbed members.  Although our sample is limited, our results suggest
that \agnfrac\ in massive galaxy clusters is not strongly correlated
with star formation at $z<1$, and that IR-AGN have a more prominent
role at $z\gtrsim1$.

\end{abstract}

% NOTE: HOW I DETERMINED REST-FRAME [3.6]um LIMITING LUMINOSITY...

%
% (1) All rest-frame 3.6\um\ flux is shifted into the [8.0] bandpass for
%      the highest-z cluster (z=1.24)
%
% (2) I took the limiting app. magnitude for the [8.0] channel in this cluster
%      and determined what flux/Hz that corresponded to.
%
% (3) Flux/Hz was converted to Luminosity/Hz assuming a distance determined
%      from $\Lambda$-CDM universe
%
% (4) Hz was removed by mult. speed of light and divide by 3.6\um\

\myfont
\section{Introduction}

Our understanding of galaxy evolution in dense cluster environments
continues to improve as observations broaden in wavelength and
redshift range. One fundamental observation is that dense environments
at low redshifts are dominated by passive, early-type galaxies that
define a narrow red sequence in an optical color-magnitude diagram
\citep{Sandage:78,Bower:92,Hogg:04}, and studies of the cluster
color-magnitude relation (CMR) show that the luminous red sequence
members in galaxy clusters have not evolved significantly since
$z\sim0.8$ \citep{Rudnick:09}.  In contrast, the less massive cluster
members continue to migrate to the red sequence as studies at $z>0.6$
show there are fewer faint red galaxies in clusters compared the field
and to lower redshift clusters \citep{DeLucia:07,Stott:07,Rudnick:09}.  Recent
observations at $z>1.4$ have now even found massive, star forming
galaxies in clusters \citep{Tran:10,Hilton:10}.

The question then remains as to what halts star formation in cluster
galaxies?  Possible environmental processes include ram-pressure
stripping \citep{Gunn:72}, tidal effects from the cluster potential
\citep{Farouki:81}, and galaxy-galaxy interactions
\citep{Richstone:76}, but none are completely effective at reproducing
the star formation histories and scaling relations observed in galaxy
clusters. One model that has proven successful is including feedback
from active galactic nuclei (AGN).  Semi-analytic models that include
AGN are able to reproduce observed mass/luminosity functions
\citep{Croton:06, Bower:06, Lagos:08}.  AGN can also affect the
intra-cluster medium (ICM) where models find that including
AGN produces much better agreement with observations of X-ray
properties of the ICM \citep{Bower:08,Puchwein:08,McCarthy:10}.

% ~~~~~~~~~~~~~~~~~~~~~~~~~~~~~ TABLE 1 ~~~~~~~~~~~~~~~~~~~~~~
\begin{deluxetable*}{lrrrrcc}
\tablecolumns{6}
\tablewidth{0pc}
\tablecaption{Cluster Properties\label{tab:clusters}}
\tablehead{ 
  \colhead{ Cluster } & 
  \colhead{ RA Dec\tablenotemark{a} } & 
  \colhead{ $z$-Range\tablenotemark{b} } &  
  \colhead{ $N_z$\tablenotemark{c} } & 
  \colhead{ $N_{\mathrm{IRAC}}$\tablenotemark{d} } & 
  \colhead{ log($L_X$)\tablenotemark{e} } & 
  \colhead{ Status\tablenotemark{f} } }
\startdata
Coma & 12  59  35.7  $\;+$27  57  34 & (0.013 - 0.033) & 348 & 262 & 43.0 & Relaxed \\
Abell 1689 & 13 11 29.5  $\;-$01 20 17 & (0.17 - 0.22) & 81 & 73 & 43.3 & Relaxed \\
MS $1358\!+\!62$ & 13 59 50.4  $\;+$62 31 03 & (0.315 - 0.342) & 171 & 133 & 43.0 & Unrelaxed \\
CL $0024\!+\!17$ & 00 26 35.7  $\;+$17 09 43 & (0.373 - 0.402) & 205 & 75 & 42.5 & Unrelaxed \\
MS $0451\!-\!03$ & 04 54 10.9  $\;-$03 01 07 & (0.52 - 0.56) & 242 & 90 & 43.3 & Relaxed \\
MS $2053\!-\!04$ & 20 56 21.3  $\;-$04 37 51 & (0.57 - 0.60) & 132 & 87 & 42.8 & Unrelaxed \\
MS $1054\!-\!03$ & 10 57 00.0  $\;-$03 37 36 & (0.80 - 0.86) & 153 & 120 & 43.2 & Unrelaxed \\
RX J$0152\!-\!13$ & 01 52 43.9  $\;-$13 57 19 & (0.81 - 0.87) & 147 & 80 & 43.3 & Unrelaxed \\
RDCS J$1252\!-\!29$ & 12 52 54.4  $\;-$29 27 18 & (1.22 - 1.25) & 38 & 29 & 42.8 & Unrelaxed \\[-2mm]
\enddata
\tablecomments{All clusters in this sample have $M_{\mathrm{vir}} \gtrsim 5 \times 10^{14} M_{\mathrm{\odot}}$.}
\tablenotetext{a}{Right ascension and Declination are for J2000.}
\tablenotetext{b}{Range of redshifts for spectroscopically confirmed
members from \citet{Saintonge:08} and \citet[RDCS
J1252-29]{Demarco:07}} 
\tablenotetext{c}{Total number of spectroscopically confirmed cluster
members. Redshifts are from Rines et al. (2003, Coma), Duc et
al. (2002, Abell 1689), Fisher et al. (1998, MS 1358), Moran et
al. (2005, CL 0024), Moran et al. (2007a, MS 0451), Tran et
al. (2005a, MS 2053), Tran et al. (2007, MS 1054), Demarco et
al. (2005, RX J0152) and Demarco et al. (2007, RDCS J1252)
respectively.}
\tablenotetext{d}{Number of cluster galaxies with detections in at
least three IRAC channels.} 
\tablenotetext{e}{Bolometric ICM X-ray luminosities (ergs s$^{-1}$)
from Holden et al. (2007, Coma, MS 1358, MS 2053, MS 1054,
RX J0152), Bardeau et al. (2007, A1689), Donahue et al. (1999, MS
0451), Zhang et al. (2005, CL 0024) and Rosati et al. (2004, RDCS
J1252). }
\tablenotetext{f}{Dynamical state of each cluster determined from
  redshift distributions and X-ray \& weak lensing profiles. Unrelaxed
  systems are those that show signs of a cluster scale merger. }
\end{deluxetable*}
% ~~~~~~~~~~~~~~~~~~~~~~~~~~~~~ TABLE 1 ~~~~~~~~~~~~~~~~~~~~~~

AGN feedback seems to be the ideal solution to resolve many
outstanding discrepancies between models and observations
\citep{Gabor:10,Fontanot:10,Teyssier:10};
%but is there observational proof that
%AGN play a pivotal role in quenching star formation and thus making
%galaxy clusters into resting homes of passive galaxies by $z\sim0$?
%%%%% replacement of {above}
however, there is not yet clear observational evidence that AGN
contribute significantly to the quenching of star formation in cluster
galaxies, making clusters into resting homes of passive galaxies.
%%%%% replacement end
Several groups using primarily X-ray observations
%supplemented by radio and infrared data
find the fractional density of X-ray selected active galactic nuclei (X-ray AGN) in
cluster environments increases with redshift
%from 0.13\% at $z\sim0.2$ to 1.00\% at $z\sim0.7$
\citep{Eastman:07,Martini:09}.
% addition begin
Similarly, using X-ray, infrared and radio selection at $z < 1.5$
\citet{Galametz:09} find that the AGN surface-density ($N$/arcmin$^2$)
is greater for clusters than in the field and that the AGN
volume-density ($N$/Mpc$^3$) for clusters increases with redshift.
% addition end
In contrast, a study of CL~$0023+04$, a large scale system of four galaxy
groups merging at $z\sim0.83$, does not find an excess of X-ray
sources relative to the field \citep{Kocevski:09}.  In general,
studies are hampered by the small number of X-ray AGN and the need to
isolate a large sample of cluster galaxies at higher redshifts.
Another important issue is that different diagnostics select different
populations of AGN \citep{Hart:09,Hickox:09,Griffith:10} and so no
single approach will be complete.  Thus while CL~$0023+04$ does not
have an excess of X-ray sources, \citet{Lubin:09} does find a
population of passive (no detectable H$\alpha$) members with [OII]
emission that may be due to AGN.

With the launch of the \textit{Spitzer Space Telescope}, mid-infrared
selection has become an efficient alternative method for identifying
AGN that complements existing X-ray, radio, and optical search
techniques.
% addition begin
Note that mid-IR color selection is particularly effective at
selecting high-Eddington AGN \citep[e.g.,][]{Hickox:09, Hopkins:09}.
% addition end
A single stellar population (SSP) has a declining
Rayleigh-Jeans tail that longward of 1$\mu$m makes these type of SEDs
blue in the mid-infrared \citep{Zhang:04} while an SED with an active
galactic nucleus has a rising power law \citep{Elvis:94} and so will
be red.  Advantages of selecting with mid-infrared wavelengths include
decreased attenuation from dust and sensitivity to AGN spanning a
broad range in redshift ($0<z\lesssim2$) \citep{Stern:05,Lacy:04}.  In
the mid-IR, red colors are due to thermal emission from heated dust,
either by radiation from stars or an active nucleus, and so dust
heated by young stellar populations, e.g. ultra-luminous infrared
galaxies (ULIRGs), can contaminate a mid-IR selected AGN sample
\citep{Donley:08}.  However, (U)LIRGs are extremely rare in massive
galaxy clusters with only one confirmed ULIRG out of thousands of
known cluster galaxies at $z<1$ \citep{Saintonge:08,Sun_Chung:10}.

%Additional multi-wavelength diagnostics can also be applied to separate star formation from AGN.

We expect that most cluster galaxies will populate a tight
distribution in mid-IR color space because massive galaxies in
clusters formed the bulk of their stars at $z>2$ and have evolved
passively since \citep{Bower:92,vanDokkum:98,Tran:07,Muzzin:08}.
However, if AGN are essential for quenching star formation in cluster
galaxies, the AGN fraction should increase with redshift because the
IR luminosity function and fraction of star forming galaxies in
massive clusters increases by a factor $\gtrsim4$ from $z\sim0$ to $z\sim1$
\citep{Saintonge:08,Bai:10}.  We can identify AGN using mid-IR colors
to determine if the IR-AGN fraction evolves with redshift and,
combined with known X-ray AGN, also study the properties of cluster AGN to
obtain a more complete picture of AGN in galaxy clusters.  Note that
selecting AGN with optical diagnostics becomes quite challenging at
$z>0.4$ because key spectral features shift into the near-infrared,
and extremely deep radio observations are significantly more time
intensive than mid-IR imaging.

Motivated by these issues, we present the first extensive mid-infrared
census of massive galaxy clusters at $0<z<1.3$ (Table~\ref{tab:clusters}).  
Our clusters are the most massive systems known
($M_{\mathrm{vir}} \gtrsim 5 \times 10^{14} M_{\mathrm{\odot}}$)
having comprehensive spectroscopic information for member
galaxies as well as deep optical photometry (see \S \ref{optical
  observations}) and so provide the most representative sample of
galaxy clusters in the range $0 \lesssim z \lesssim 1.3$. This paper is organized
as follows: in \S 2 we discuss the reductions of the
\textit{Spitzer} mid-infrared and optical data for each cluster. In \S
3 we describe how we select AGN and results for each cluster. In \S 4
we measure the IR-AGN fraction as a function of redshift and discuss
properties of their host galaxies, and we present our conclusions in
\S 5. All magnitudes are in the AB system unless otherwise noted.  We
assume a $\Lambda$CDM cosmology throughout with \Om = 0.27 , \Olam =
0.73 and $h$ = 0.71

\section{Data and Reductions}

\subsection{Spitzer IRAC}

%Data were
Mid-infrared observations were taken with the IRAC instrument \citep{Fazio:04} on board the
\textit{Spitzer Space Telescope} \citep{Werner:04} and are publicly
available on the Spitzer archive.
%IRAC takes broadband photometry
IRAC observes
in four mid-infrared channels centered at 3.6, 4.5, 5.8 and 8.0
\um. These channels have transmission functions such that all emission
between 3.1 and 9.4 \um\ will be detected. Furthermore, the edges of
the channels are steep, so that as an emission feature becomes
redshifted out of one bandpass it will promptly shift into the
adjacent one. This ensures that spectral features (such as PAH
emission from dusty star formation) can be traced fairly easily over a
wide range in redshift.

For most of the cluster fields (Table~\ref{tab:clusters}), multiple
IRAC observations were conducted at different times. In the interest
of increasing depth we make use of all available data. This also
frequently served to increase the area of coverage allowing more
cluster galaxies to be detected.

\subsubsection{Mosaicking}

Each observational campaign is composed of a series of dithered images
referred to as Basic Calibrated Data (BCD). The depth of a final
mosaic depends on the combination of exposure time and number of
frames per sky position. BCDs are single-frame images that have been
reduced and flux-calibrated on a basic level by the \textit{Spitzer}
pipeline. We performed further processing and mosaicking using
MOPEX\citep[18.3 rev 1,][]{Makovoz:06}, a set of reduction and
analysis tools designed by the \textit{Spitzer Science Center}. The
software includes modules that flag and remove spurious detections
that are not accounted for in the automatic pipeline reduction. Prior
to mosaicking, images were inspected for artifacts such as muxbleed
and column pull-up/down and were mitigated using the $cosmetic$ module
from MOPEX.

Overlapping is another necessary step in the image reduction process
available as a package in MOPEX. Within a set of dithered frames, the
individual backgrounds may vary to the point that the background of a
final mosaic shows a checkered pattern/gradient. Performing overlap
matches the backgrounds of all input frames and so removes this
effect. After overlapping, BCDs are then stacked. Weight maps were
also obtained for each mosaic as coverage varied between campaigns.

\subsubsection{Photometry and Completeness}
\label{phot}

Fixed-aperture photometry was carried out on each mosaic using
SExtractor 2.5.0 \citep{Bertin:96}. In order to analyze cluster
galaxies at varying $z$ equally, apertures were chosen at constant
proper sizes according to cluster redshifts (except for Coma, see \S
\ref{coma}). Ideally, apertures should be small so as to isolate the
central engine and reduce possible contamination from, e.g. extended
star forming regions in the host galaxy. However, the IRAC point
spread functions (PSFs) are \twid 1.66, 1.72, 1.88 and 1.98'' in
diameter for channels 1$-$4 respectively, and fluxes determined from
apertures comparable to the IRAC pixel size ($\sim1.22''$ px$^{-1}$)
are subject to sampling errors. These caveats constrain the minimum
size of a reasonable aperture. Ultimately, we choose a radius of \twid
12.6 kpc, which corresponds to an aperture diameter of 3'' for the
most distant cluster (see Table \ref{completeness table} 
for all aperture sizes). 
% addition begin
We note that a 12.6 kpc aperture is much
larger than would be ``ideal'' for this type of a study; however, we feel
that accepting this limitation is preferred over varying the physical sizes of
apertures, which would likely introduce a selection bias.
% addition end

Aperture corrections were determined from IRAC calibration stars
\citep{Reach:05} as discussed in \citet{Ashby:09}
%  addition begin
, except for Coma where theoretical aperture corrections for extended
sources were used. Although some galaxies in our other low-$z$
clusters are resolved, the discrepancy between extended- and
point-source aperture corrections is not enough to affect our results.
% addition end
Further discussion regarding these corrections can be found in the
IRAC Instrument Handbook. 
IRAC fluxes are calibrated based on $24''$ (diameter)
apertures and an appropriate correction needs to be applied to
photometry using apertures of different sizes. Aperture corrections
are defined as the difference between the magnitude of a point-sources
from a $24''$ aperture and the magnitude from the aperture of
interest. We determined aperture corrections from five standard stars
(HD-165459, 1812095, KF06T1, NPM1p66.0578, NPM1p67.0536) using the
average as the final value. Because Galactic extinction is negligible
at these wavelengths ($\lesssim 0.01$ mag) the corrections are
less than the measurement uncertainties, we do not correct the IRAC
magnitudes; however, we do correct for Galactic extinction in the
optical filters.
%IRAC magnitudes were not corrected for
%Galactic extinction since at these wavelengths corrections would be
%smaller than the measurement uncertainties ($\lesssim 0.01$ mag).

\begin{figure*}[t]
\epsfig{ file=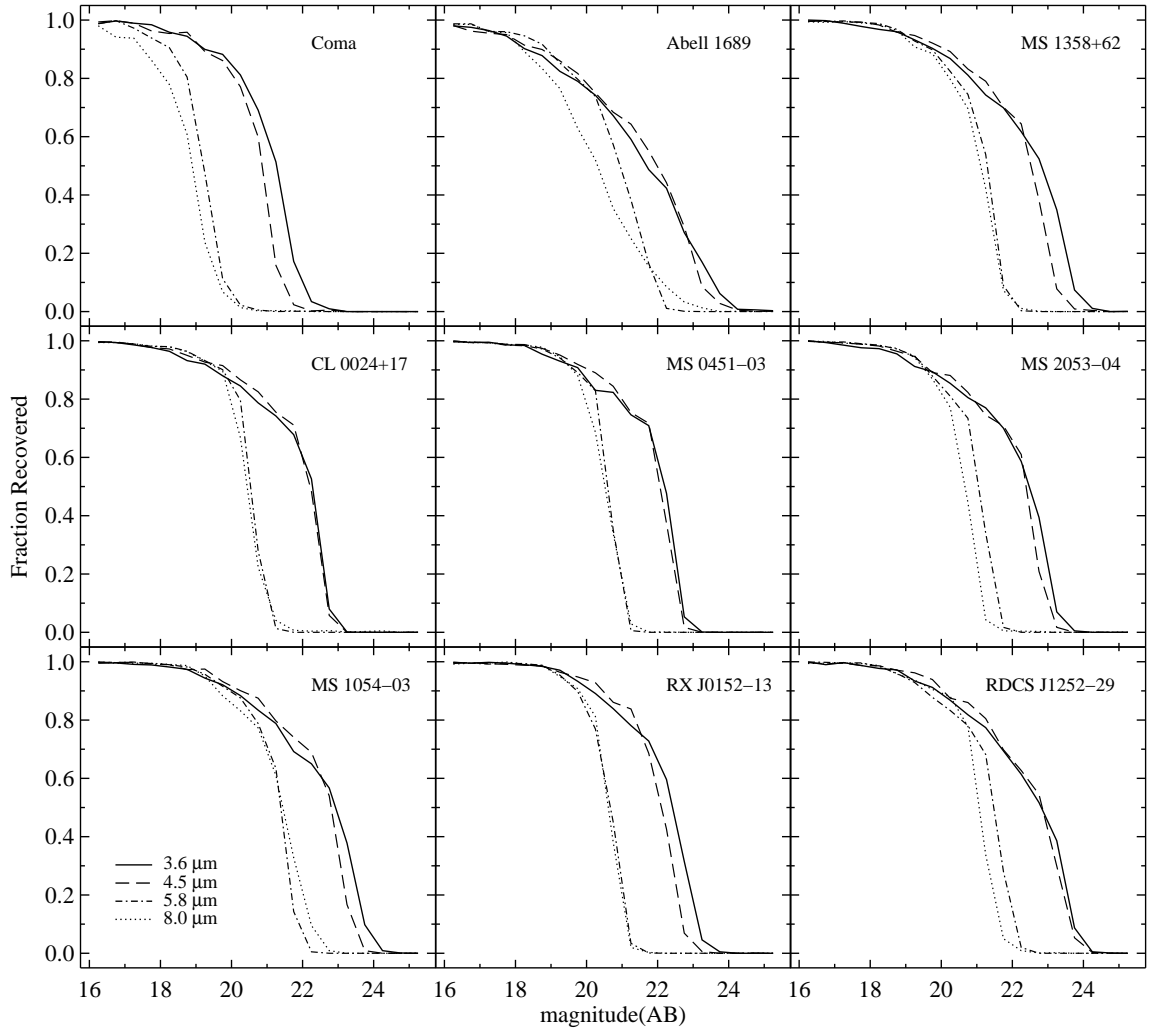 }
\caption{\figtxt Completeness functions for each cluster field showing
the fraction of recovered sources versus their input magnitude. For
each cluster, 1000 generated galaxies were randomly distributed in 0.5
magnitude bins in a 1.5\x1.5 Mpc$^2$ box centered near the cluster
core. Sources that were extracted with magnitudes brighter than their
input value were rejected as blendings. The 3.6 and 4.5\um\ channels
exhibit a more gradual decline in completeness due to source confusion
from greater crowding \citep[see][]{Barmby:08} which increases the
likelihood of blendings. } 
\label{completeness figure}
\end{figure*}

Completeness was measured using the \textit{gallist} and
\textit{mkobjects} modules in IRAF\footnotemark[1]. For each mosaic,
1000 artificial galaxies in half-magnitude bins between $16 \le m \le
25$ were distributed randomly. Source extraction was then carried out
for these fake sources with identical parameters as used for the real
sources. Fake sources that were extracted with magnitudes brighter
than their input value were discarded as blendings. In order to get a
sense of the completeness within each cluster, sources were added in
1.5\x1.5 Mpc boxes centered on the core of each cluster. Measurements
of completeness are shown in Figure \ref{completeness figure} and
Table \ref{completeness table}.  \footnotetext[1]{ IRAF is distributed
by the National Optical Astronomy Observatory (NOAO), which is
operated by the Association of Universities for Research in Astronomy,
Inc., under cooperative agreement with the National Science
Foundation. }

%In performing source matching, we start with the coordinates of a galaxy
%from optical images. We then search for the nearest source in each
%IRAC catalog within a 2'' radius. In some cases (particularly for the
%high-$z$ clusters) blending of cluster galaxies occured in the IRAC
%imaging. Deblending was performed using the SExtractor parameters
%DEBLEND\_NTHRESH$=64$ (for all channels) and DEBLEND\_MINCONT$=0$ and
%$0.005$ (for channels 1,2 and 3,4 respectively). This approach was
%very effective at deblending galaxies in the cores clusters. This may
%not be surprising since the IRAC PSFs are roughly the same size as the
%cluster galaxies. Nevertheless, upon visual inspection roughly 1-2
%pairs/groups of galaxies were still blended in the $z > 0.8$ clusters.
In performing source matching, we start with the coordinates of a
galaxy from optical images. We then search for the nearest source in
each IRAC catalog within a 2" radius, i.e. slightly larger than the
IRAC PSF. In a few cases, e.g. in the cores of the high redshift
clusters, there is some blending of IRAC sources.  We deblend and
separate sources using the SExtractor parameters DEBLEND NTHRESH=64
(for all channels) and DEBLEND MINCONT=0 and 0.005 (for channels 1,2
and 3,4 respectively).  Visual inspection confirms that these
parameters are effective at deblending sources with only 2
pairs/groups of galaxies still blended in the $z>0.8$ clusters; only
one of these has an IR-AGN signature (see \S 3.2.8).

% ~~~~~~~~~~~~~~~~~~~~~~~~~~~~~ FIGURE 1 ~~~~~~~~~~~~~~~~~~~~~~
% ~~~~~~~~~~~~~~~~~~~~~~~~~~~~~ FIGURE 1 ~~~~~~~~~~~~~~~~~~~~~~

\subsection{Optical Photometry and Spectroscopy}
\label{optical observations}

% ~~~~~~~~~~~~~~~~~~~~~~~~~~~~~ TABLE 2 ~~~~~~~~~~~~~~~~~~~~~~
\begin{deluxetable}{lrrrrr}
\tablecolumns{6}
\tablewidth{0pc}
\tablecaption{IRAC Photometry\label{completeness table}}
\tablehead{
  \colhead{Cluster} & 
  \colhead{Aperture\tablenotemark{a}} & 
  \colhead{3.6\um\ \tablenotemark{b}} & 
  \colhead{4.5\um\ \tablenotemark{b}} &
  \colhead{5.8\um\ \tablenotemark{b}} & 
  \colhead{8.0\um\ \tablenotemark{b}} }
\startdata
Coma & 15'' & 20.3 & 20.1 & 18.8 & 18.1 \\
1689 & 7.7'' & 19.6 & 19.9 & 19.7 & 19.0 \\
1358 & 5.3'' & 20.8 & 21.1 & 20.4 & 20.2 \\
0024 & 4.8'' & 20.6 & 20.9 & 20.2 & 20.0 \\
0451 & 4.0'' & 20.9 & 21.0 & 20.3 & 20.0 \\
2053 & 3.8'' & 20.8 & 20.9 & 20.3 & 20.1 \\
1054 & 3.3'' & 21.1 & 21.2 & 20.7 & 20.5 \\
0152 & 3.3'' & 21.1 & 21.4 & 20.1 & 20.3 \\
1252 & 3.0'' & 21.0 & 21.3 & 20.6 & 20.6 \\
\enddata
\tablenotetext{a}{Fixed circular aperture corresponding to physical
diameter of $\sim25$ kpc at cluster redshift except for Coma where
the aperture corresponds to 6 kpc
%due to resolved sources
; see \S3.2 for more details. } 
\tablenotetext{b}{AB magnitude corresponding to 80\% completeness
limit; see \S\ref{phot} for more details. }
\end{deluxetable}
% ~~~~~~~~~~~~~~~~~~~~~~~~~~~~~ TABLE 2 ~~~~~~~~~~~~~~~~~~~~~~

Catalogs of optical and near infrared photometry as well as optical
spectroscopy were obtained from multiple sources. Observed photometry
was converted to rest-frame values using KCorrect v0.2.1, an extension
of $kcorrect$ \citep{Blanton:07} developed for Python by Taro
Sato. Extensive spectroscopic catalogs were used to confirm membership
for each cluster. General properties of each galaxy cluster are shown
in Table 2. To summarize:

% ~~~~~~~~~~~~~~~~~~~~~~~~~~~~~ FIGURE 2 ~~~~~~~~~~~~~~~~~~~~~~
\begin{figure*}[t]
\epsfig{ file=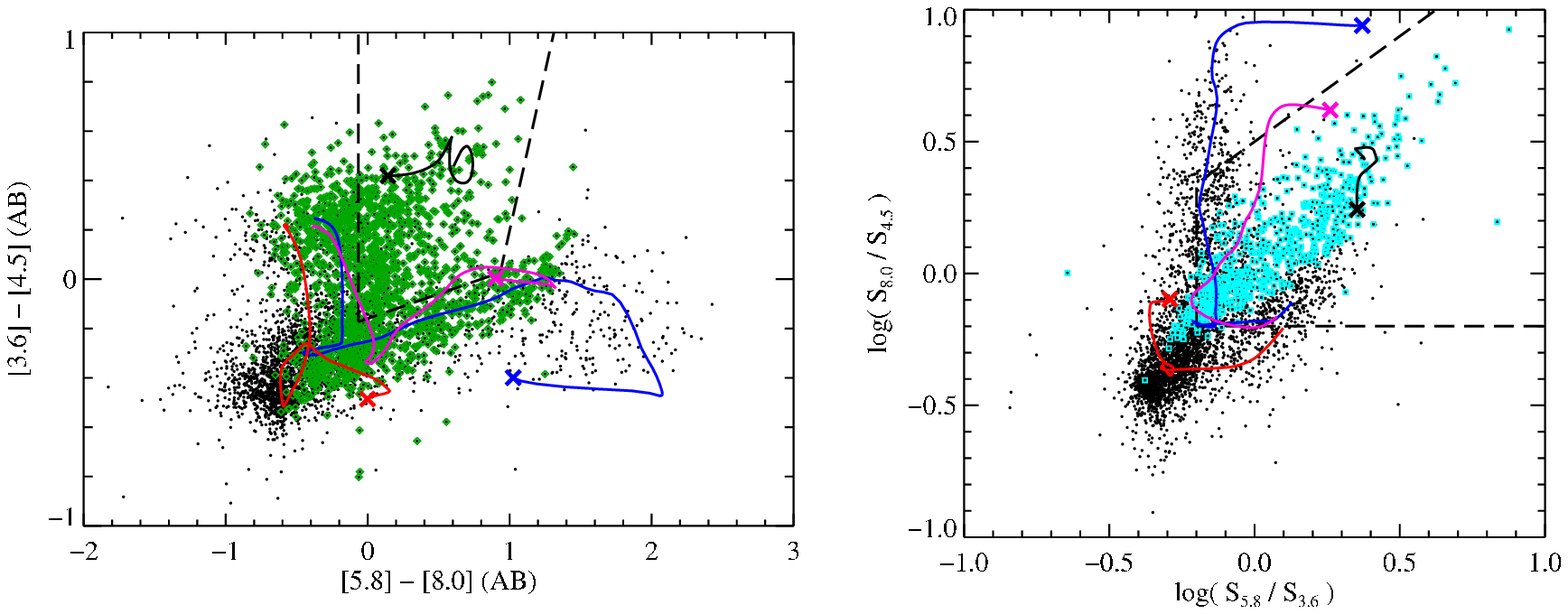 }
\caption{\figtxt Combined IRAC color-color plots as defined by
\citet[left]{Stern:05} and \citet[right]{Lacy:04} derived from 3''
apertures. Plotted are all detected sources in the IRAC mosaics for
each cluster field. Curves in both panels 
are redshift-tracks for M82 (a local starbursting galaxy; blue), VCC
1003 (a local passively evolving galaxy; red), Mrk231 (Seyfert 1 AGN;
black) and a Seyfert 2 template (pink); tracks begin at $z=0$ marked
with Xs and go to $z=2$ \citep{Devriendt:99, Polletta:07}. The areas
enclosed by the dashed lines are empirically defined regions designed
to select galaxies dominated by emission from an AGN. Sources that
fall in the wedge on the \textit{left} are overlayed with cyan squares
on the \textit{right} whereas sources in the wedge on the
\textit{right} are overlayed with green diamonds on the
\textit{left}. Only $\sim$33\% of Lacy IR-AGN are selected as Stern IR-AGN,
whereas $\sim$89\% of Stern IR-AGN are selected as Lacy IR-AGN. Furthermore,
the Stern criteria determine an AGN fraction of 15\%, whereas using
the Lacy criteria it is 40\%. We adopt the 
Stern IR selection for our analysis as it seems to suffer less
contamination from non-AGN sources. See \citet{Barmby:06} and
\citet{Donley:08} for a more in depth analysis of mid-IR color
evolution in galaxies with and without AGN. } 
\label{wedgies}
\end{figure*}
% ~~~~~~~~~~~~~~~~~~~~~~~~~~~~~ FIGURE 2 ~~~~~~~~~~~~~~~~~~~~~~

\begin{itemize}

\item \textit{Coma}: Photometry was taken from \citet{Mobasher:01}
which consists of measurements in the B and R bands. Completeness was
assessed to be 22.5 mag and 21 mag in $B$ and $R$
respectively. Spectroscopy for the Coma cluster was taken as part of
the Cluster and Infall Nearby Survey
\citep[CAIRNS:][]{Rines:03}. Galaxies targeted in this survey were
selected from digitized images of the POSS I 103aE (red) plates which
are complete down to $E$=15-16.

\item \textit{Abell 1689}: Data for Abell 1689 \citep{Duc:02} were
taken as follow-up to the photometric observations conducted by
\citet{Dye:01}. Photometry was acquired for the $BVR$ bands and is
complete to 23.0, 22.7 and 22.7 mag respectively. Spectra were
obtained for most(\twid 75\%) of the photometric cluster members at $R
\le 17.75$ mag which drops to \twid40\% at $R \le 19.5$ mag.

\item \textit{MS 1358+62}: Observations of MS 1358 taken from
\citet{Fisher:98} including photometry in the $V$ and $R$ bands and
spectroscopy. Spectroscopic completeness was determined to be
$>\!\!80\%$ at $R \le 21$ mag when compared to photometric
observations which were complete to $R \sim 23.5$ mag.

\item \textit{CL 0024+17 and MS 0451-03}: Data for CL 0024 and MS
0451, including photometry from the HST WFPC2 instrument, are
discussed in detail in \citet{Treu:03} and
\citet[2007]{Moran:05}. Photometry was measured to be complete to $I
\sim 25$ (Vega mags)\footnotemark[2] for CL 0024 and spectroscopic
completeness was found to be $>65$\% at $I<21.1$ and $I<22.0$ mag for
CL 0024 and MS 0451 respectively. Additional ground-based photometry
was also obtained in the $BVRIJK_s$ bands reaching $3 \sigma$ depths
of 27.8, 26.9, 26.6, 25.9, 21.6 \& 19.7 mag for CL 0024 and 28.1,
27.0, 27.3, 25.9, \& 20.2 mag for MS 0451 \citep{Moran2:07}.
\footnotetext[2]{ Here the $I$-band refers to the F814W filter from
the WFPC2 instrument}

\item \textit{MS 2053-04 and MS 1054-03}: Spectroscopy for MS 2053 is
detailed in \citet{Tran:05a} and completeness determinations were
assessed according to sampling and success rates. The success rate is
defined as the number of spectroscopic redshifts obtained divided by
the number of spectroscopic targets. Spectroscopic completeness was
determined to be \twid 70\% at $m_{814} < 22$ mag. Similar methods
were applied for MS 1054 \citep{Tran:07} which found completeness to
be $>\!\!75\%$ at $m_{814} \le 21.2$ mag. Photometry for both MS 2053
and MS 1054 were acquired from the $HST$/WFPC2 F606W and F814W
filters.

\item \textit{RX J0152-13 and RDCS J1252-29}: Photometry for RX J0152
\citep{Blakeslee:06} was obtained from the ACS instrument onboard the
$HST$ in the F625W, F775W and F850LP bandpasses.  Incompleteness for
these observations begins to set in at \twid 23.5, 22.5 and 22.
\citet{Demarco:05} determined spectroscopic membership for RX J0152
confirming 102 cluster galaxies out of 262 targets. Observations of
RDCS 1252 is outlined in \citet{Demarco:07}. Photometry was taken in
the $BVR i_{775} z_{850}J K_s$ filters reaching $5 \sigma$ limiting
magnitudes of 26.5 \& 26 mag in the $J$ \& $K_s$ filters
respectively. The spectroscopic success rate for RDCS J1252 was found
to show a rapid decline at $K_s = 21.5$ mag, dropping from 85\% to
20\%.

\end{itemize}

\section{Results}
\label{results}

\subsection{IRAC Color Selection of AGN}

Mid-IR emission from an AGN is widely accepted as thermal continuum
from circumnuclear dust \citep{Andreani:03,Kuraszkiewicz:03}. As
radiation from the accretion disk bombards the surrounding dust, it is
heated to temperatures in the range of $\sim 20-1000$ K \citep[below
dust sublimation at $T \sim 2000$ K,][]{Sanders:89}. IRAC colors alone
can be an effective method for separating star forming galaxies from
those hosting AGN at redshifts up to $z\sim2$
\citep{Lacy:04,Stern:05}. It is worth noting that $\gtrsim 50\%$ of a galaxy's
mid-IR emission must originate from the nuclear component
\citep{Hickox:07, Hopkins:09, Atlee:11} in order to
be selected by the criteria of \citet{Stern:05}. Thus, due to various limitations in
measuring IRAC fluxes in our sample, we are only able to select
galaxies that are dominated by AGN emission in the mid-IR.

The optical to mid-IR spectral energy
distribution (SED) of AGN are typically well characterized by a rising
power law with a few notable PAH features, thus causing these objects
to appear increasingly red in the IRAC window
\citep{Polletta:07}. Consequently, AGN tend to populate a separate
region in mid-IR color space. Furthermore, studies have shown that
various types of Seyferts have very similar mid-IR properties
\citep[and references therein]{Gandhi:09}. Radiation at these wavelengths is
relatively insensitive to extinction and thus gives a reliable measure
of reprocessed emission from the central engine. However, it is worth
noting that highly obscured sources ($A_V \gtrsim 30$) may be pushed
outside the IRAC selection wedge \citep[see Fig. 1 of][]{Hickox:07}.

% ~~~~~~~~~~~~~~~~~~~~~~~~~~~~~ FIGURE 3 ~~~~~~~~~~~~~~~~~~~~~~
\begin{figure*}[t]
\centering
\epsfig{ file=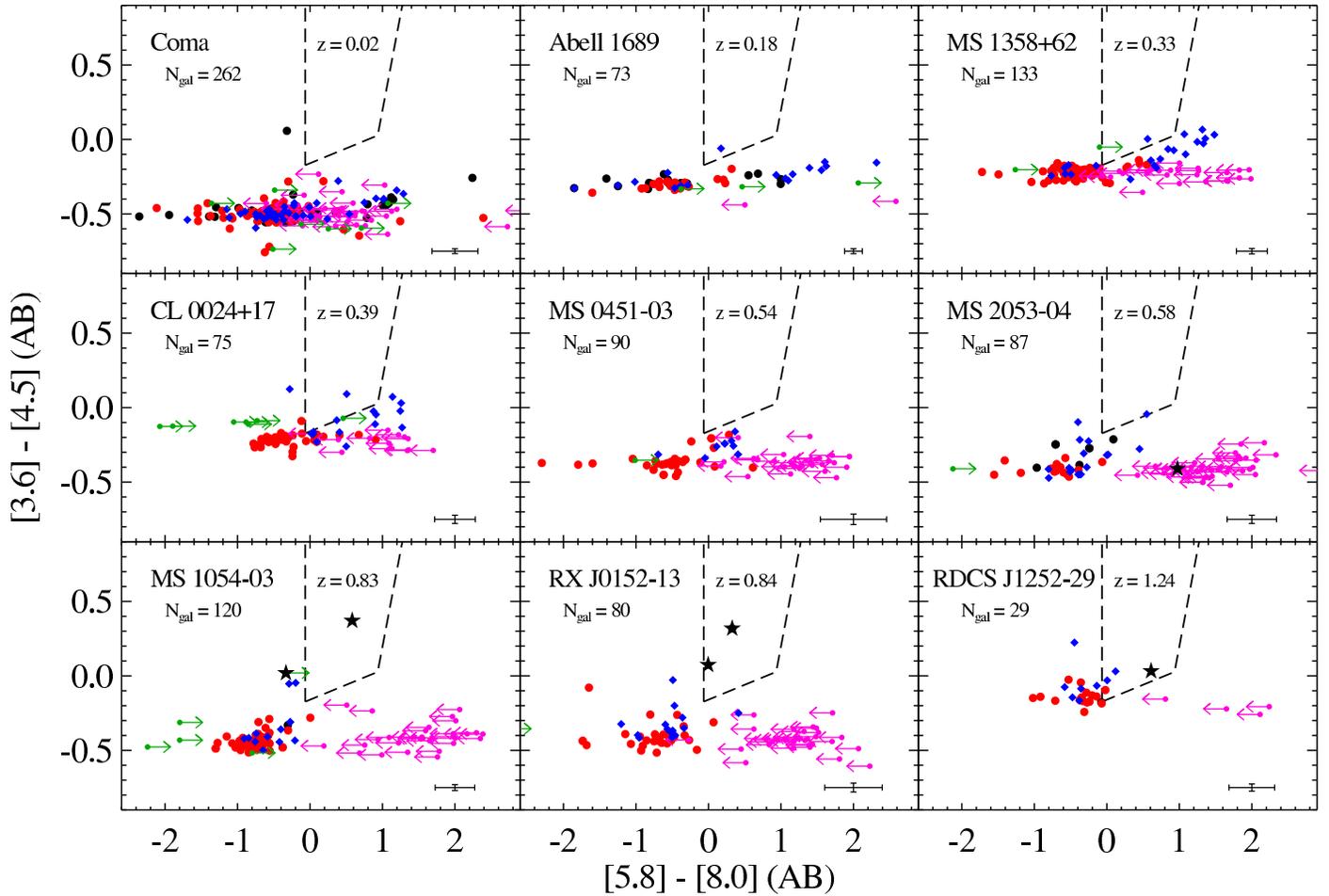 , width=\linewidth }

\caption{\figtxt IRAC color-color plots used to select active galactic
nuclei \citep[IR-AGN;][]{Stern:05} for the galaxy clusters in our
study. Cluster redshift and number of confirmed members are shown in
each panel. Data points correspond to morphologically classified
early-type members (red circles), late-types (blue diamonds) and
unclassified/mergers (black circles). Upper limits are determined for
galaxies lacking detections in a particular channel (green and pink
arrows, see \S \ref{results}). X-ray sources are indicated as
stars. Mean uncertainties are shown in the lower-right corner of each
panel. Early-type galaxies with blue IRAC colors populate the
lower-left region in each plot (the ``passive cloud'') whereas IR-AGN
populate the area enclosed by the dashed lines. Our cluster IR-AGN
are predominantly hosted by late-type galaxies. } 
\label{plots}
\end{figure*}
% ~~~~~~~~~~~~~~~~~~~~~~~~~~~ FIGURE 3 ~~~~~~~~~~~~~~~~~~~~~~

Figure \ref{wedgies} shows the empirical color selection criteria from
both \citet[left]{Stern:05} and \citet[right]{Lacy:04} for all IRAC
sources detected in our nine cluster fields (including the cluster
galaxies). Included in Figure \ref{wedgies} are color tracks of
template galaxy SEDs from \citet{Devriendt:99}: M82 (local
starbursting galaxy; blue), VCC 1003 (local passively evolving galaxy;
red), Mrk231 (Seyfert 1 AGN; black), and a Seyfert 2 template (pink).
Tracks begin at $z=0$ marked with Xs and go to $z=2$
\citep{Polletta:07}.  We refer the reader to \citet{Barmby:06} and
\citet{Donley:08} for a more detailed discussion of the mid-IR color
evolution for galaxies including limitations of mid-IR selection.

% before
%The IR-AGN fractions determined for all the sources in our images are
%15\% and 40\% for the Stern and Lacy criteria respectively.
% before
To compare the selection methods of \citet{Lacy:04} and
\citet{Stern:05} we determine IR-AGN fractions for all galaxies (field
and cluster) detected in all four bands of our IRAC imaging. We obtain
fractions of 40\% and 15\% for the Lacy and Stern criteria respectively.
In Figure
\ref{wedgies}, the green diamonds in the Stern plot correspond to
sources that are selected as AGN using the Lacy criteria, whereas cyan
squares in the Lacy plot correspond to sources that are selected as
AGN using the Stern criteria. Approximately 89\% of the Stern-selected
AGN are Lacy-selected AGN; however, the reverse shows that only
\twid33\% of all Lacy-selected AGN are also selected based on the
Stern criteria.  The higher IR-AGN fraction measured using the Lacy
criteria is not surprising given that the track for M82 falls in the
Lacy wedge.  Because the Lacy criteria do not exclude starburst
galaxies as effectively as the Stern criteria, we adopt the Stern
criteria throughout the rest of this paper.

\subsection{Individual Clusters}

IRAC color plots for the nine massive galaxy clusters are shown in
Figure \ref{plots}. Only galaxies that have been spectroscopically
confirmed as members with $\ge\!3\sigma$ detections in at least 3 IRAC
channels are shown. Data points indicate morphologically classified
elliptical/S0 galaxies (red circles), late-type galaxies (blue
diamonds), unclassified/merger (black circles), sources with no
detection in channel 4 only (pink arrows) and no detection in channel
3 only (green arrows). For sources that lacked detections in a single
bandpass, upper/lower limits were determined by assuming the 80\%
completeness magnitude for the respective bandpass (Table
\ref{completeness table}). Applying these limits mostly reveals a
fainter population of passive galaxies.  Considering that channels 1
\& 2 (shorter wavelength) probe to fainter magnitudes than channels 3
\& 4 (longer wavelength), galaxies in the ``passive cloud'' (with
declining mid-IR SEDs) tend not to be detected at longer wavelengths
while galaxies with IR-AGN (with increasing mid-IR spectra) are more
likely to be deteced in all four channels.  This is why no potential
candidate AGN are identified by our limit determinations. We use
optically-determined coordinates for cluster galaxies to locate 
their IRAC counterparts. Using a matching radius of 2'' (6'' for Coma)
we find the rate of detecting a false positive to be $<1\%$.

\subsubsection{Coma}
\label{coma}

The Coma cluster is one of the richest and most closely studied galaxy
clusters and is known to be dominated by passively evolving systems
with early-type morphologies \citep{Michard:08}. IRAC imaging for this
cluster in all four channels covers roughly a 51.1'\x 62.5' region
centered on NGC 4874, a field of view that includes 348
spectroscopically confirmed cluster galaxies from the CAIRNS
\citep{Rines:03}. Determining photometry with \textit{Spitzer} data
for resolved galaxies is not as straightforward as for the mostly
unresolved galaxies in the distant clusters (see the IRAC Instrument
Handbook for details).  We decide against using flexible-aperture
photometry from SExtractor because of the differently sized apertures
used for the same galaxy across the IRAC channels. Instead, we use a
constant $15''$ diameter aperture (6 kpc at Coma's mean redshift) and
apply aperture corrections to the extended objects as detailed in the
IRAC Instrument Handbook.

Not surprisingly, we find that an overwhelming majority of Coma
galaxies occupy the ``passive cloud'' in IRAC color space and well
below the AGN wedge.  The dispersion in the [5.8]$-$[8.0] color is
likely due to PAH features from star formation at 6.2\um\ and 7.7\um\
being detected in IRAC's 8.0\um\ bandpass; note that Coma's proximity
means we detect even the faintest members.  There is one member with a
significantly redder [3.6]$-$[4.5] color that is a disk galaxy
viewed at an intermediate angle, but it is not an IR-AGN from the
\citet{Stern:05} criteria and is not classified as an optical AGN in
the recent survey of Coma by \citet{Mahajan:10}.

\subsubsection{Abell 1689} % photometry from Duc et al. was kcorrected

Spectroscopically confirmed members and their photometry are from
\citet{Duc:02}. The scatter in [5.8]$-$[8.0] color among members
can be attributed to PAH features at 3.3, 6.2 and 7.7\um\ in dusty
star forming galaxies where the latter two features both shift into
the 8.0\um\ channel. However, the 3.3\um\ PAH feature shifts to the
boundary between channels 1 \& 2. We find one candidate IR-AGN for
this low redshift cluster that has also been classified as a Seyfert 1
AGN based on optical spectroscopy \citep{Duc:02}. This prior study of
Abell 1689 included mid-IR measurements from ISOCAM and the authors
concluded that dusty star formation in this cluster is responsible for
the vast majority of the observed mid-IR emission. Our results support
this conclusion as we find no other members with infrared AGN
signatures.

\subsubsection{MS 1358+62} % photometry from Fisher et al. was kcorrect

Photometry and spectroscopic information of MS 1358 members is from
\citet{Fisher:98}. At $z=0.33$, the 3.3 and 6.2\um\ PAH features shift
into IRAC channels 2 \& 4 respectively and colors along both axes are
redder.  This explains the dispersion seen in Figure \ref{plots} for
late-type galaxies that are likely to be star forming. The
spectroscopic study of this cluster by \citet{Fisher:98} revealed a
number of emission line galaxies (ELGs). Not surprisingly, nearly all
of the galaxies that depart from the ``passive cloud'' are also
ELGs. Of the galaxies in this cluster, we find one candidate AGN that
is hosted by a late-type spiral. \citet{Martini:09} find no X-ray AGN in
this cluster with $L$(2-8 keV) $\geq 10^{43}$ ergs s$^{-1}$. The one
galaxy that we select as an AGN is detected as an ELG located roughly
860 kpc from the cluster center.

\subsubsection{CL 0024+17} % photometry from moran et al. 2007a was kcorrected

Spectroscopy and photometry for CL 0024 is from \citet{Moran1:07}.
star forming members with PAH emission are subject to the same effects
as described for MS 1358 and thus produce a similar scatter in mid-IR
color-color space. The scatter here makes it difficult to discern the
nature of the galaxies that are found near the boundary of the
wedge. We do find two galaxies in the AGN-wedge (one appears to be
sitting on the boundary). For the galaxy on the boundary, it is
probable that star formation is producing its colors, though it is not
ruled out as hosting an IR-weak AGN. The other candidate AGN we find
shows colors consistent with a power-law spectrum placing it well
inside the IRAC wedge. \citet{Zhang:05} have conducted an X-ray
observation of this cluster locating a handful of point sources at
$L_X > 10^{42}$ ergs s$^{-1}$. However, none of the X-ray point sources
overlap with any cluster galaxies from the spectroscopic catalog,
i.e. these X-ray sources are in the field.

\begin{figure*}
\centering

\epsfig{ file=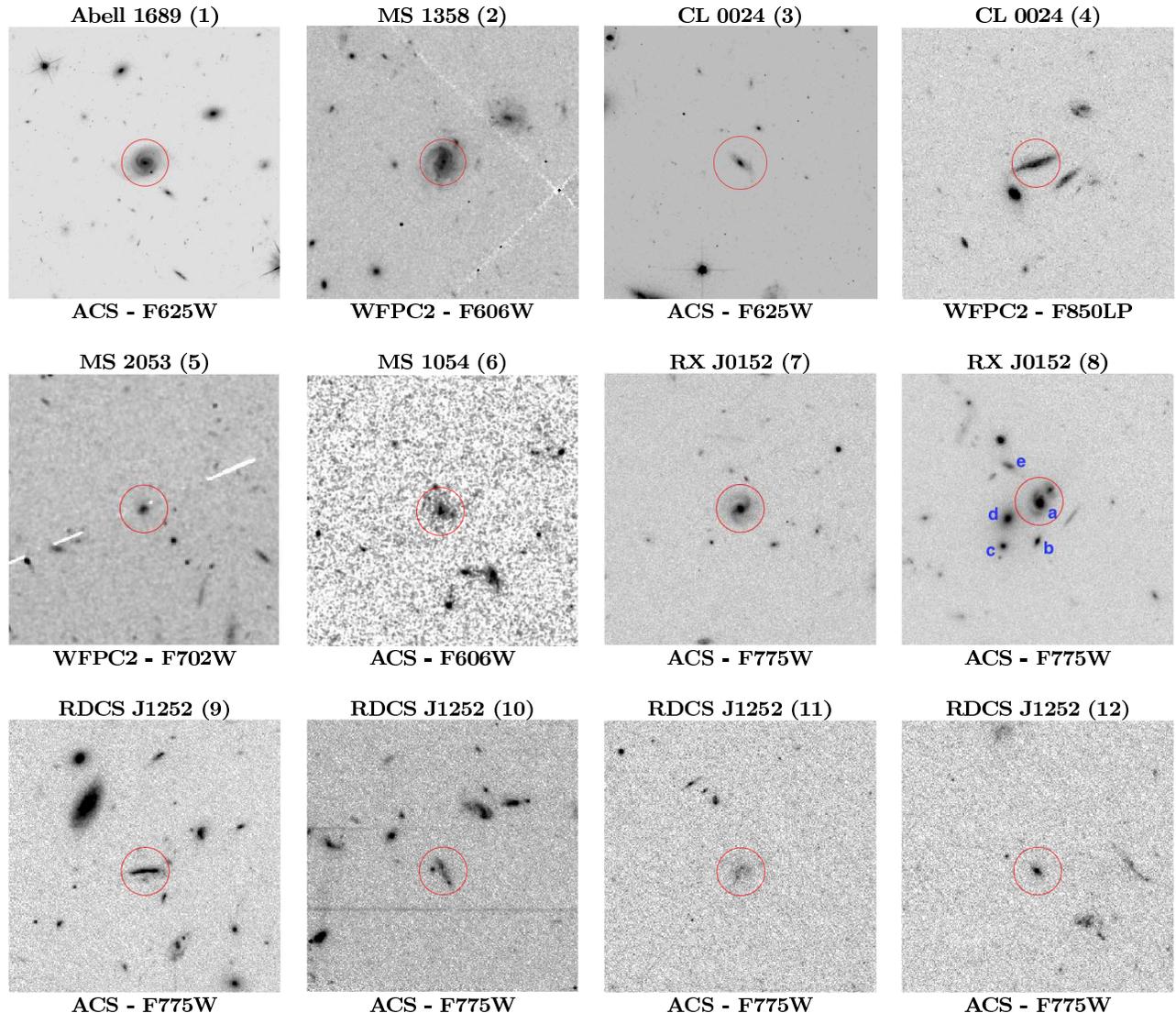 }

\caption{ Thumbnails of our sample of infrared-selected AGN; images are from the \textit{Hubble Space Telescope}. Numbers in parentheses correspond to the ID from Table \ref{agn_stats}. All images are \twid 150$\times$150 kpc and are oriented North-up, East-left. Red circles correspond to the aperture of constant size ($r \approx 12.6$ kpc) used to perform photometry on the IRAC images. See \S \ref{rxj0152} for an explanation of the labels in panel 8. }
\label{thumbs}
\end{figure*}

CL 0024 is known to have numerous substructures as traced by three
techniques: weak-lensing map \citep{Kneib:03}, X-ray shock fronts
\citep{Zhang:05} and a Dressler-Shectman test
\citep{Moran1:07,Dressler:88}. This implies that many galaxies are in
groups that have been recently accreted (or are in the process) into
the main cluster. Furthermore, the redshift distribution of galaxies
shows a bimodality that suggests a recent merger along the line of
sight with a large galaxy group \citep[2002]{Czoske:01}. The one
candidate AGN that is well inside the wedge has two close neighbors
(confirmed spectroscopically) and is located \twid0.84 Mpc from the
cluster center (see Figure \ref{thumbs}). Redshifts of these galaxies
are consistent with the main cluster. The candidate IR-AGN that we
find near the edge of the IRAC-wedge is also within the main cluster;
it is closer to the cluster core (\twid340 kpc) but is $>$80 kpc from
the nearest neighboring galaxy.

\subsubsection{MS 0451-03} % photometry from moran et al. 2007a was kcorrected

Spectroscopy and photometry for MS 0451 are also from
\citet{Moran1:07}.  At $z=0.54$, the 6.2\um\ PAH feature shifts nearly
outside the IRAC window. This reduces the scatter in mid-IR color due
to star formation for members. X-ray data have shown that the
distribution of the cluster ICM is predominantly elliptical
\citep{Donahue:03}, and the redshift distribution of cluster galaxies
is broadly Gaussian \citep{Moran1:07}. This indicates that MS 0451 is
predominantly virialized with no substantial infalling galaxies.  We
detect no IR-AGN and no cluster galaxies are detected as X-ray sources
based on a $Chandra$ survey \citep{Molnar:02}, further indicating a
lack of strong nuclear activity in MS 0451.

% ~~~~~~~~~~~~~~~~~~~~~~~ FIGURE 4 ~~~~~~~~~~~~~~~~~~~~~
% ~~~~~~~~~~~~~~~~~~~~~~~ FIGURE 4 ~~~~~~~~~~~~~~~~~~~~~

\subsubsection{MS 2053-04}

Spectroscopic and photometric information of galaxies in this cluster
come from \citet{Tran:05a}.  Detailed spectroscopic and gravitational
lensing studies of MS 2053-04 \citep{Verdugo:07,Tran:05a} show that it
is a merger of two structures with 113 and 36 confirmed members
respectively. Galaxies in the smaller structure (2053-B) have similar
properties to field galaxies not associated with the cluster.  This,
coupled with the high fraction of star forming members (\twid $44$\%),
suggests that MS 2053 has yet to completely virialize.

Observations with \textit{Spitzer}/MIPS also find that a fairly high
fraction (\twid18\%) of cluster members are detected at 24\um\
\citep{Saintonge:08}, and the large population of star forming members
clearly separates from the passive members in mid-IR color.  Given the
high level of activity in MS 2053, it is surpising then that it only
has one weak IR-AGN candidate.  
%$Chandra$ observations do identify one
%X-ray point source associated with the cluster \citep{Eastman:07}, but
%this X-ray member's [3.6]-[4.5] color is enough to exclude it as
%an IR-selected AGN.
%
% addition begin
Using archival $Chandra$ data, \citet{Eastman:07} find five X-ray
sources at the redshift of MS 2053 with $L_{X,H} > 10^{42}$
ergs/s. However, three are not classified as cluster members due to
their distances from the cluster center ($r > r_{200}$) and another is the BCG
which is thought to be contaminated by X-ray emission from the
ICM. This leaves only one cluster X-ray AGN which has a [3.6]$-$[4.5]
color that is bluer than the IR-AGN selection region (Figure
\ref{plots}).
% addition end

% ~~~~~~~~~~~~~~~~~~~~~~~~~~~ TABLE 3 ~~~~~~~~~~~~~~~~~~~~~
\begin{deluxetable*}{rrrrrrrrrr}
%\rotate
\tablecolumns{10}
\tablewidth{0pc}
\tablecaption{IRAC-selected Cluster AGN\label{agn_stats}}
\tablehead{
  \colhead{ID} &
  \colhead{Cluster} &
  \colhead{RA (J2000)} &
  \colhead{Dec (J2000)} &
  \colhead{$z$\tablenotemark{a}} &
  \colhead{$R_{proj}$ \tablenotemark{b}} &
  \colhead{$M_{3.6}$\tablenotemark{c}} &
  \colhead{HR\tablenotemark{d}} &
  \colhead{log$(L_{X})$\tablenotemark{e}} &
  \colhead{Morphology\tablenotemark{f}}}
\startdata
1 & Abell 1689 & 13 11 35.5 & $-$01 20 12.8 & 0.2000 (1) & 0.30 & -21.66$\pm0.04$ & \nodata & \nodata & Sc \\
2 & MS 1358 & 13 59 24.0 & +62 31 08.0 & 0.3236 (2) & 0.86 & -21.64$\pm0.03$ & \nodata & \nodata & Sc \\
3 & CL 0024 & 00 26 40.0 & +17 09 41.8 & 0.3955 (3) & 0.33 & -21.36$\pm0.05$ & \nodata & \nodata & Sa+b \\
4 & CL 0024 & 00 26 33.7 & +17 12 19.8 & 0.3964 (3)  & 0.84 & -21.64$\pm0.05$ & \nodata & \nodata & Sc+d \\
5 & MS 2053 & 20 56 21.0 & $-$04 37 22.8 & 0.5763 (4)  & 0.19 & -21.48$\pm0.10$ & \nodata & \nodata & S0/a \\
6 & MS 1054 & 10 57 02.7 & $-$03 39 43.6 & 0.8319 (5)  & 1.02 & -24.12$\pm0.07$ & 0.03$\pm0.18$ & 43.23 & Irr \\
7 & RX J0152 & 01 52 43.8 & $-$13 59 01.3 & 0.8201 (5)  & 0.78 & -23.61$\pm0.08$ & -0.62$\pm0.05$ & 44.18 & Sb \\
8 & RX J0152 & 01 52 39.8 & $-$13 57 40.7 & 0.8300 (5) & 0.48 & -24.88$\pm0.07$  & -0.09$\pm0.07$ & 44.52 & merger \\
9 & RDCS 1252 & 12 52 55.6 & $-$29 27 09.7 & 1.2274 (6) & 0.15 & -21.85$\pm0.12$ & \nodata & \nodata & Irr \\
10 & RDCS 1252 & 12 52 57.4 & $-$29 27 32.0 & 1.2322 (6) & 0.35 & -22.28$\pm0.12$ & \nodata & \nodata & Irr \\
11 & RDCS 1252 & 12 52 49.8 & $-$29 27 54.7 & 1.2382 (6) & 0.59 & -22.95$\pm0.12$ & 0.20$\pm0.31$ & 43.15 & Irr \\
12 & RDCS 1252 & 12 52 49.7 & $-$29 28 03.7 & 1.2382 (6) & 0.65 & -22.61$\pm0.12$ & \nodata & \nodata & S0 \\
\enddata
\tablenotetext{a}{Spectroscopic redshift: (1) Duc et al. 2002 ; (2)
Fisher et al. 1998 ; (3) Moran et al. 2007a ; (4)
Tran et al. 2005a ; (5) Holden et al. 2007 ; (6) Demarco et al. 2007.}
\tablenotetext{b}{Projected distance from the cluster center in Mpc.}
\tablenotetext{c}{Rest-frame 3.6\um\ absolute magnitude.}
\tablenotetext{d}{X-ray hardness ratio from \citet{Martel:07} for AGN
with X-ray detections.}
\tablenotetext{e}{Hard X-ray (2-10 keV) luminosity in ergs s$^{-1}$ from
Martel et al. 2007.} 
\tablenotetext{f}{Morphology references: \citet[Coma]{Michard:08},
\citet[Abell 1689]{Duc:02}, \citet[MS 1358]{Fabricant:00}, \citet[CL
0024]{Moran1:07}, \citet[MS 0451]{Moran1:07}, \citet[MS
2053]{Tran:05b}, \citet[MS 1054]{Blakeslee:06}, \citet[RX
J0152]{Blakeslee:06}  and \citet[RDCS J1252]{Demarco:07}.} 
\end{deluxetable*}
% ~~~~~~~~~~~~~~~~~~~~~~~~~~~ TABLE 3 ~~~~~~~~~~~~~~~~~~~~~

\subsubsection{MS 1054-03}

Spectroscopically confirmed members and photometry are taken from
\citet{Tran:07}. One cluster X-ray source was not included in the
photometric catalog, but optical and X-ray data for this galaxy are
available from \citet{Martel:07}. Weak-lensing and X-ray analyses of
MS 1054 show a clumpy nature to the dark matter and ICM profiles
\citep{Jee:05b}. The presence of such substructure indicates that the
cluster experienced a merger and has yet to fully virialized.

Earlier studies of MS 1054 reveal that it contains two members hosting
X-ray AGN \citep{Johnson:03} and 8 radio sources that can be powered by
AGN or star formation \citep{Best:02}.  We find only one infrared AGN
at a distance $\gtrsim$1 Mpc from the cluster center that is also
detected as both an X-ray and radio source.
%The second X-ray source is $>$1 Mpc from the cluster BCG and lies outside the IRAC AGN wedge.
%
% addition begin
The second X-ray source lies near the edge of the IRAC footprints and
is not detected at 5.8\um . Though this does not allow it to be
identified as IR-AGN, it is not ruled out based on its [3.6]$-$[4.5]
color (Figure \ref{plots}).
% addition end

\subsubsection{RX J0152-13} %  photometry from blakeslee et al. 2006 was kcorrected
\label{rxj0152}

Spectroscopic membership of cluster galaxies is from
\citet{Demarco:05} and photometry from \citet{Blakeslee:06}. This
cluster shows signs of having gone through a large-scale merger event
recently as indicated by its X-ray emission, luminosity distribution
and weak-lensing profile \citep{Jee:05a}.

We detect two IR-AGN in this cluster, both of which are classified as
X-ray QSOs \citep{Martel:07}; no other cluster galaxies are detected
as X-ray sources. One of the IR+X-ray AGN is about 800 kpc from the
cluster core while the second appears to be in a merging system.  The
latter detection is associated with five cluster galaxies within a
projected radius of 30 kpc (Figure \ref{thumbs}, neighboring cluster
galaxies are labeled). Due to their proximity and the IRAC PSF, these
galaxies are blended into one mid-IR source where galaxy $a$ is the
closest (\twid 0.7'') to the centroid of the mid-IR
emission. Redshifts in this quintet are 0.867, 0.864, 0.834, 0.832 and 0.836 for
galaxies $a$, $b$, $c$, $d$ and $e$ respectively. 
%This indicates that galaxies $a$ \& $b$ are likely in a bound system
%and galaxies $c$, $d$ \& $e$ are projected members.
%
% addition begin
Based on these redshifts, recessional velocities ($w.r.t.$ to galaxy
$c$) are 3743, 3408, 0, -231 and 231 km/s for galaxies $a$, $b$, $c$,
$d$ and $e$ respectively. Since galaxies $a$ \& $b$ have velocities
within 400 km/s and galaxies $c$, $d$ \& $e$ have velocities within
500 km/s (typical of galaxy groups) we suspect $a$ \& $b$ are a bound
system and $c$, $d$ \& $e$ are another bound system.
% addition end
% addition begin
Note, however, that a chance alignment of two groups such as this is
\twid 0.1\% likely to occur at random assuming a spherical cluster geometry
with $R \approx 1$ Mpc. Of course, due to the complex morphology and
substructure in RX J0152 \citep{Jee:05a} this probability may increase.
% addition end

\subsubsection{RDCS J1252-29}

At $z=1.24$, RDCS 1252 is the most distant cluster in our sample yet
has a virial mass similar to those of the lower redshift clusters as well as other properties
\citep{Rosati:04}. Close inspection of the ICM in the cluster core
reveals the presence of a shock front, signaling a recent merger with
a cluster sub-clump. A detailed spectroscopic follow-up
\citep{Demarco:07} verifies a merger of two groups that have yet to
viralize, and weak lensing shows that the centroid of the dark matter
mass profile is offset from the optical/X-ray centroid by \twid8''
\citep{Lombardi:05}. Yet despite its young dynamical age, RDCS 1252
already has a population of luminous early-type galaxies that show
little sign of ongoing star formation.

At this redshift, emission from stellar populations begins to encroach
into the IRAC window. This effect can be seen in Figure \ref{plots} as
an upward shift in the the ``passive cloud'' relative to the lower-$z$
clusters. This unfortunately brings the ``passive cloud'' closer to
the AGN-wedge, possibly introducing contamination. Of the twenty nine
RDCS 1252 members shown in Figure \ref{plots}, four are inside the
AGN-wedge, of which one has been previously identified as an X-ray AGN
\citep{Martel:07,Demarco:07}. Three of the four candidate IR-AGN are
hosted by morphologically irregular galaxies (Figure \ref{thumbs})
that are likely to be gravitationally disrupted because of the
large-scale cluster merger or by galaxy-galaxy mergers (\textit{HST}
image from \citet{Blakeslee:03} and \citet{Demarco:07}).

\section{Discussion}

\subsection{Cluster IR-AGN Properties}

Recent studies of field galaxies find that infrared-selected AGN share
similar properties \citep{Hickox:09,Griffith:10}: their host galaxies
typically have late-type morphologies and tend to have blue optical
colors.  The properties of our 12 cluster IR-AGN agree with these
earlier studies. We find the majority of our IR-selected AGN (10/12)
are in late-type galaxies (Table \ref{agn_stats}) that are blue
(Fig.~\ref{cmd}) and so have recent/ongoing star formation. 
The remaining two cluster IR-AGN are hosted by S0 galaxies that lie
on/near the IR-AGN boundary (Fig.~\ref{plots}) and these members are
consistent with being spiral galaxies transitioning to early-type
systems in the cluster environment \citep{Dressler:97,Moran1:07}.  

% addition begin
It is worth noting that optical light from host galaxies may be
contaminated by emission from an AGN, thus their colors could be
biased bluewards. Our bluest AGN host, for example, is likely
contaminated by the central engine since it is such a strong outlier. 
Figure \ref{thumbs} shows ACS \& WFPC-2 images
of our IR-selected AGN, nearly all at rest-frame blue wavelengths.
Upon careful visual inspection we find that the majority of these
galaxies have extended morphologies and are not strongly dominated by
a central point source. 
Furthermore, \citet{Hickox:09} calculated color contamination to be $\lesssim
0.3$ mag in $^{0.1}$($u-r$)\symbolfootnote[1]{This color is computed
  by blueshifting the SDSS $u$ \& $r$ filters by $z=0.1$}. When
considering that the optical filters we use
($B-V$) are closer together in wavelength space and that AGN generally
contribute more flux at bluer wavelengths, we expect color
contamination in our sample to be less than what \citet{Hickox:09}
find. Therefore, although point source contamination may be impacting some of our
sample, we conclude that color contamination of our IR-AGN
hosts is not significant enough to bias our results.
% addition end

% ~~~~~~~~~~~~~~~~~~~~~~~~~~~~~ FIGURE 5 ~~~~~~~~~~~~~~~~~~~~~~
\begin{figure}[t] %\hspace{-1.4cm}
\centering
\epsfig{file=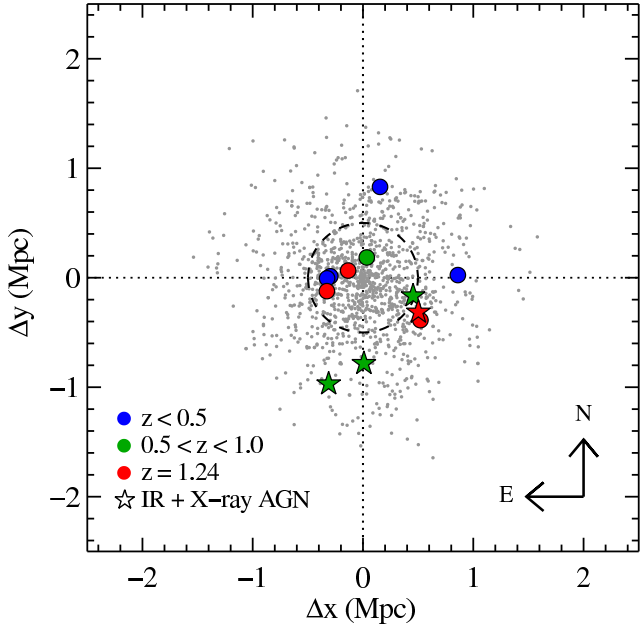 , width=0.95\linewidth}
\caption{\figtxt Combined sky-plot for each cluster showing the
  projected locations of galaxies with respect to their cluster
  centers. IRAC-selected AGN are shown as circles (or stars for X-ray sources)
and the dashrd fircle corresponds to a physical radius of 0.5
Mpc. Colors correspond to three bins is redshift, $z < 0.5$ (blue),
$0.5 < z < 1.0$ (green) and $z=1.24$ (red). A Kolmogorov-Smirnov test
shows that the radial distribution of the IR-AGN is $>$99\% likely to
come from the same parent population as the normal cluster galaxies
(gray points). }
\label{skyplot}
\end{figure}
% ~~~~~~~~~~~~~~~~~~~~~~~~~~~~~ FIGURE 5 ~~~~~~~~~~~~~~~~~~~~~~

We find that a third (4/12) of our cluster IR-AGN are also known X-ray
sources \citep{Martel:07,Johnson:03}, a fraction that is nominally
consistent with results from the AGN and Galaxy Evolution Survey
(AGES) where $\sim50$\% of IR-AGN are also X-ray sources
\citep{Hickox:09}. However, it is worth mentioning here that X-ray
flux limits are not the same among these studies, and so agreement on
this ratio is not necessarily implied. The X-ray sources are also the
four most luminous 3.6\um\ IR-AGN (rest-frame; Figure \ref{cmd});
however, note that the most luminous cluster IR+X-ray AGN is the
blended source in RX J0152 (see \S3.2.8).

Figure \ref{skyplot} shows the projected distances of the 12 IR-AGN
relative to confirmed members in all nine galaxy clusters. We find the
radial distribution of the IR-AGN is drawn from the same parent
population as the cluster galaxies with $>99$\% confidence using a
Kolmogorov-Smirnov test. However, the four IR+X-ray AGN are all
outside the cluster cores at $R_{proj}\gtrsim 0.5$ Mpc. This result is
consistent with \citet{Eastman:07} and \citet{Atlee:11} who find that
X-ray sources in galaxy clusters are not strongly centrally concentrated.
%but at odds with \citet{Martini:07} and \citet{Atlee:11} who find that
%cluster X-ray AGN are more centrally concentrated than normal cluster galaxies.
%
%However, the four IR+X-ray AGN are all
%outside the cluster cores at $R_{proj}\gtrsim 0.5$ Mpc which is consistent
%with \citet{Eastman:07} who find that X-ray sources in galaxy clusters
%are not strongly centrally concentrated.
Our observations may suggest that IR+X-ray AGN represent a different
population than IR-only AGN; however, we are limited by the size of
our sample and so cannot further postulate on the uniqueness of these
AGN based on their spatial distribution and 3.6\um\ luminosities.
% addition begin
%Our observations might suggest that IR+X-ray AGN represent a different
%population than IR-only AGN; however, since we are limited by the size
%of our sample it is difficult to postulate on the uniquness of these
%active galactic nuclei (based on $R_{proj}$ and 3.6\um\
%luminosity). Recent studies of local ($z<0.05$) Seyferts agree that
%there is a significant correlation between the mid-IR continuum and
%X-ray luminosities of the nucleus
%\citep{Krabbe:01,Lutz:04,Horst:08}. Unfortunately, aquiring a
%measurement of the nuclear mid-IR continuum from IRAC
%photometry is not possible at the redshifts of our sample. This
%restricts us even further from speculating about our AGN sample.
% addition end

% ~~~~~~~~~~~~~~~~~~~~~~~~~~~~~ FIGURE 6 ~~~~~~~~~~~~~~~~~~~~~~
\begin{figure} %\hspace{-1.4cm}
\centering
\epsfig{file=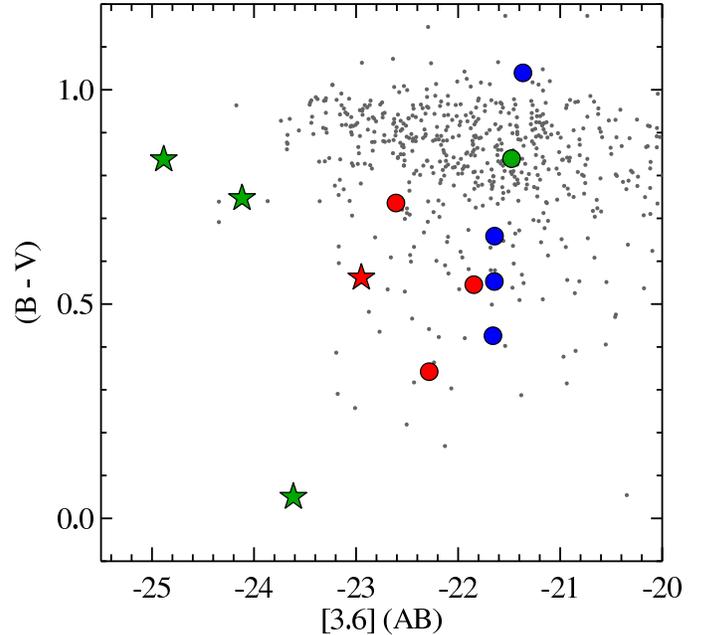 , width=1.0\linewidth}
\caption{\figtxt Rest-frame color-magnitude relation (optical color
vs. 3.6$\mu$m absolute magnitude); symbols are the same as in figure
\ref{skyplot}. Host galaxies of IR-AGN tend to have blue optical
colors, thus these IR-AGN hosts have recent/ongoing
star formation. Also, the four most luminous cluster IR-AGN are also
X-ray sources. }
\label{cmd}
\end{figure}
% ~~~~~~~~~~~~~~~~~~~~~~~~~~~~~ FIGURE 6 ~~~~~~~~~~~~~~~~~~~~~~

%Although we are limited by
%our small sample, our analysis indicates that IR-AGN in clusters
%differ from the IR+X-ray AGN population as the latter tend to be
%outside the cluster core and are the most luminous IR sources.  For
%example, the IR+X-ray AGN may be gas-rich galaxies on their first
%entry into the cluster where tidal interactions brings fresh gas to
%their black holes and triggers a powerful (radiative) X-ray phase.

\subsection{Infrared-AGN Fractions}

To measure the fraction of IR-AGN in our cluster sample and test for
evolution, we separate our sample into three redshift bins:
low redshift ($<0.5$), intermediate redshift ($0.5<z<1.0$), and a high
redshift point at $z=1.24$ (RDCS 1252) containing 543, 377 and 29 IRAC
detected members respectively.  To ensure robustness, we consider two
different galaxy samples selected optically and in the mid-IR.  We
also take into account the varying spatial coverage of the IRAC
mosaics and set the maximum field of view with the Coma cluster where
the IRAC footprint only includes galaxies within $R_{proj}\sim760$ kpc
of the cluster center.  In the higher redshift clusters, we therefore
exclude members that are at $R_{proj}>760$ kpc from their cluster
center.

\begin{figure}[t]
\epsfig{ file=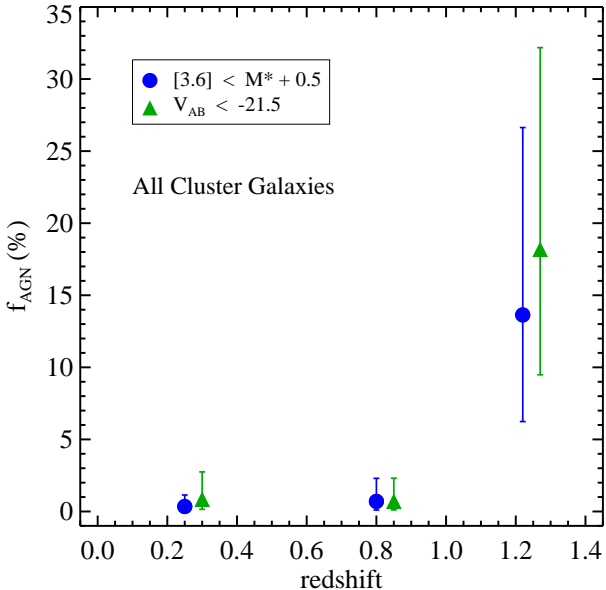 , width=1.3\linewidth }
\caption{ \figtxt Cluster IR-AGN fraction as a function of redshift
for optically selected members brighter than $V_{AB}=-21.5$ mag (green
triangles) and mid-IR selected members brighter 
than (\Mstar$\,+\, 0.5)$ (blue circles).  We consider three redshift
bins: low redshift ($z<0.5$), intermediate redshift ($0.5<z<1.0$), and
a high redshift point at $z=1.24$ (RDCS 1252) that contain 543, 377,
and 29 IRAC-detected members respectively.  The IR-AGN fraction is
uniformly $\lesssim3$\% at $z<1$ and only measurably higher in RDCS
1252 at $z=1.24$. Error bars represent 1$\sigma$ Poisson uncertainties derived
using statistics from \citet{Gehrels:86}. }
\label{fractions all}
\end{figure}

Our first cluster galaxy sample is composed of optically-selected members
brighter than $V = -21.5$ (this corresponds to where the $V$-magnitude
distribution turns over for RDCS 1252, our most distant cluster); this
yields 118, 141 and 22 galaxies in our three redshift bins.  Note we
do not to correct for passive evolution given that the host galaxies
of the IR-AGN tend to have blue optical colors, i.e. are likely
star forming systems.  The IR-AGN fraction for this optically-selected
sample is $\sim1$\% for both redshift bins at $z<1$ and is only
measurably non-zero at $z=1.24$ with \agnfrac=$18.2^{+14}_{-8.7}$\% (Table
\ref{fractions table}; Fig.~\ref{fractions all}). All errors in
\agnfrac\ are asymmetric 1$\sigma$ Poisson uncertainties as determined
by \citet{Gehrels:86} for small number samples.

%Because the evolution of the 3.6\um\ luminosity function is
%well-characterized by a system of passively evolving galaxies that
%formed at $z > 1.5$ \citep{Muzzin:08}, our second cluster sample is
%composed of rest-frame 3.6\um\ selected members, i.e. essentially a
%stellar mass cut. 
%
% replacement begin {above par.}
Our second cluster galaxy sample is composed of members selected based
on rest-frame 3.6\um\ luminosity. Because of the evolution of the
3.6\um\ luminosity function is well-characterized by passively
evolving galaxies that formed at $z>1.5$ \citep{Muzzin:08}, this
luminosity selection is effectively a stellar mass cut. 
% replacement end {above par.}
To isolate comparable samples of cluster members
over our redshift range, we combine values of $M^{\star}(z)[3.6\mu$m]
\citep[the characteristic turning point in the Schechter luminosity
function;][]{Schechter:76}) from \citet{Muzzin:08} with the 80\%
limiting magnitude for our most distant cluster and thus select
members brighter than rest-frame (\Mstar$(z)[3.6\mu$m$]+0.5$).  We find
that the cluster IR-AGN fraction is again uniformly $\sim1$\% at
$z<1$ and only measurably non-zero at $z=1.24$ with
\agnfrac=$13.6^{+13}_{-7.4}$\% (RDCS 1252; Table \ref{fractions table};
Fig.~\ref{fractions all}).

% ~~~~~~~~~~~~~~~~~~~~~~~~~~~~~ FIGURE 7 ~~~~~~~~~~~~~~~~~~~~~~
% ~~~~~~~~~~~~~~~~~~~~~~~~~~~~~ FIGURE 7 ~~~~~~~~~~~~~~~~~~~~~~

\begin{figure}[t]
\epsfig{ file=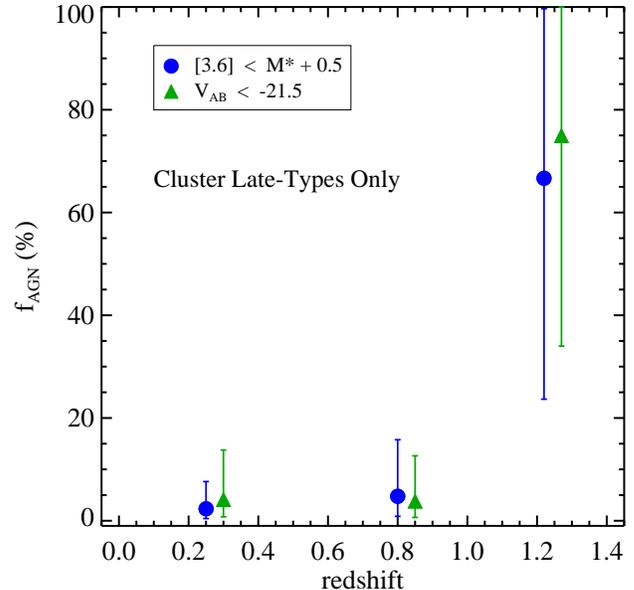 , width=1.3\linewidth }
\caption{ \figtxt Same as Fig. \ref{fractions all} but only
considering late-type galaxies (i.e. excluding E/S0 galaxies).   The
IR-AGN fraction remains $\sim5$\% at $z<1$ and is higher only at $z=1.24$. }
\label{fractions late}
\end{figure}

Thus far we have included all cluster galaxies regardless of
morphology in determining \agnfrac, but this may introduce a bias
given that: 1) our cluster IR-AGN are predominantly hosted by
late-type galaxies and 2) the morphological mix in clusters evolves
with redshift \citep{Dressler:97, Fasano:00,Postman:05,Capak:07}.  In
Fig.~\ref{fractions late} we now exclude all morphologically
classified E/S0 members and measure a higher \agnfrac\ at all
redshifts (Table \ref{fractions table}).  However, \agnfrac\ remains
$\lesssim5$\% at $z<1$ in both of our selected galaxy samples.  Only
in the most distant cluster (RDCS 1252) does \agnfrac\ for
late-type\footnote{Here we mean all members except for E/S0s.} members
increase to $\sim70$\%.

Although the number of cluster IR-AGN is small, we stress that our
analysis is based on a sample of $\sim1500$ spectroscopically
confirmed cluster galaxies at $0<z<1.3$, thus we place a strong upper
limit on the IR-AGN fraction of $\lesssim3$\% for all members in
massive clusters at $z<1$.  One caveat to consider is that while the
IR color selection does identify $\sim90$\% of broad-line AGN, it
misses $\sim60$\% of narrow-line AGN \citep{Stern:05} and so we may be underestimating
\agnfrac.  However, strongly starbursting galaxies may also
contaminate our IR-AGN sample \citep[up to $\sim50$\%;][]{Donley:08},
and accounting for these actually decreases the IR-AGN fraction.
Addressing these two competing effects is beyond the scope of our
current analysis.

% ~~~~~~~~~~~~~~~~~~~~~~~~~~~~~ FIGURE 8 ~~~~~~~~~~~~~~~~~~~~~~
% ~~~~~~~~~~~~~~~~~~~~~~~~~~~~~ FIGURE 8 ~~~~~~~~~~~~~~~~~~~~~~

% ~~~~~~~~~~~~~~~~~~~~~~~~~~~~~ TABLE 4 ~~~~~~~~~~~~~~~~~~~~~~
\begin{deluxetable}{l|rcrr}
\tablecolumns{6}
\tablewidth{0pc}
\tablecaption{IR-AGN Fractions\label{fractions table}}
\tablehead{
  \colhead{Selection\tablenotemark{a}}   &   
  \colhead{$z$-bin}   &   
  \colhead{$N_{\tiny{\mathrm{AGN}}}$}   &   
  \colhead{$N_{tot}$}   &   
  \colhead{\agnfrac\tablenotemark{b} } }
\startdata
All Members        &   $(z<0.5)$   &   1   &   291   &   0.3$^{+0.8}_{-0.3}$ \% \\
M$_{3.6}<$ \Mstar$+0.5$ &   $(0.5<z<1.0)$  &   1   &   143   &   0.7$^{+1.6}_{-0.6}$ \% \\
	    	           &   $z=1.24$   &   3   &   22   &   13.6$^{+13}_{-7.4}$ \% \\[1mm]
\hline \\[-2.5mm]
All Members        &   $(z<0.5)$   &   1   &   118   &   0.8$^{+1.9}_{-0.7}$ \% \\
$V_{\rm{AB}} < -21.5$       &   $(0.5<z<1.0)$   &   1   &   141   &   0.7$^{+1.6}_{-0.6}$ \% \\
		           &   $z=1.24$   &   4   &   22   &   18.2$^{+14}_{-8.7}$ \% \\[1mm]
\hline \\[-2.5mm]
Late-Types Only            &   $(z<0.5)$   &   1   &   43   &   2.3$^{+5.3}_{-1.9}$ \% \\
M$_{3.6}<$ \Mstar$+0.5$ &   $(0.5<z<1.0)$   &   1   &   21   &   4.7$^{+11}_{-3.9}$ \% \\
		           &   $z=1.24$   &  2   &   3   &   67$^{+33}_{-43}$ \% \\[1mm]
\hline \\[-2.5mm]
Late-Types Only            &   $(z<0.5)$   &   1   &   24   &   4.1$^{+9.6}_{-3.4}$ \% \\
$V_{\rm{AB}} < -21.5$     &   $(0.5<z<1.0)$   &   1   &   26   &   3.8$^{+8.8}_{-3.2}$ \% \\
		           &   $z=1.24$   &   3   &   4   &   75$^{+25}_{-41}$ \%\\[-2mm]
%
%  the errors of +33% and +25% were originally
%  +88% and +73%, but i changed it to be 100% at max
%
\enddata
\tablenotetext{a}{Members are selected using a luminosity limit
in rest-frame 3.6\um\ or rest-frame $V_{AB}$.}
\tablenotetext{b}{Uncertainties in \agnfrac\ represent 1$\sigma$
  Poisson errors determined from \citet{Gehrels:86}.}
\end{deluxetable}
% ~~~~~~~~~~~~~~~~~~~~~~~~~~~~~ TABLE 4 ~~~~~~~~~~~~~~~~~~~~~~

The low IR-AGN fraction of $\sim1$\% in our massive clusters at
$z<1$ is consistent with \citet{Martini:09} who estimate an X-ray AGN
fraction of $0.13-1.00$\% in clusters at $\bar{z}\sim0.2$ to
$\bar{z}\sim0.7$ and with \citet{Hickox:09} who find the IR-AGN
fraction in AGES is comparable to the X-ray AGN fraction at $0.25 < z
< 0.8$.  However, we
cannot say for certain that there is strong evolution in the cluster
IR-AGN fraction with redshift given our small numbers.  This is in
contrast to the observed increase in the fraction of (dusty)
star forming members in these same clusters \citep{Saintonge:08}, thus
the bulk of their $24\mu$m flux is due to star formation and not AGN.
Our single galaxy cluster at $z>1$ does suggest that IR-AGN have a
more prominent role at this epoch, but we recognize that 1) RDCS 1252
may be unusually active and 2) the IRAC color selection starts to
suffer contamination from passive galaxies at these redshifts.  Given
their rarity, a larger survey of IR-AGN in massive galaxy clusters,
particularly at $z>1$, is needed to robustly identify any evolution in
IR-AGN with redshift.

Another interesting comparison we can study is the variation in
\agnfrac\ with environment.
%A difference in the frequency of AGN in low to high density
%environments would provide insight into various properties of AGN
%evolution \citep[fueling mechanisms, duty cycles, etc;
%see][]{Hopkins:06}.
%
%addition begin
Using magnitude cuts similar to our $V \le -21.5$ limit, the
 \agnfrac\ in the Bo\"otes field sample from the AGES ($0.25 < z < 0.8$) is \twid 2\%
 (R. Hickox, private communication). This is well within our
upper 1$\sigma$ uncertainty (Table \ref{fractions table})
at similar redshifts, thus there is no clear variation in
\agnfrac\ based on local density. The possibility of AGN
playing a more influential role at $z>1$ still remains.
% addition end

\section{Conclusions}

We present the first census of mid-infrared selected active galactic
nuclei (IR-AGN) in massive galaxy clusters ($M_{\mathrm{vir}} \gtrsim 5
\times 10^{14} M_{\mathrm{\odot}}$) at $0<z<1.3$ by combining
archival \textit{Spitzer}/IRAC imaging with extensive optical
spectroscopic catalogs (public and private) and deep optical
photometry of $\sim1500$ confirmed members in nine clusters.  Our
clusters are selected to be the most massive well-studied systems
currently known.  Using
the four IRAC channels (3.6, 4.5, 5.8 and 8.0\um) and established
mid-IR color selection techniques \citep{Stern:05,Lacy:04}, we
identify 949 members that are detected ($>3\sigma$) in at least three
of the four IRAC channels and isolate 12 that host dust-enshrouded
AGN.  Similar to IR-selected AGN in recent field studies
\citep{Hickox:09,Griffith:10}, the host cluster galaxies tend to be
late-type members with blue optical colors that indicate
recent/ongoing star formation. The IR-AGN have the same radial
distribution as the cluster members, but the four most IR-luminous AGN
lie outside of their cluster cores ($R_{proj}>0.5$ Mpc)
%
%   OLD TEXT:   and are also known X-ray sources, i.e. very bright IR+X-ray AGN are
%   OLD TEXT:   not centrally concentrated in their clusters \citep{Martini:09}.
%
%  revision begin
and are also known X-ray sources. This suggests that very bright
IR+X-ray AGN are not centrally concentrated in their clusters,
consistent with the results for bright X-ray sources by \citet{Martini:09}.
% revision end
Our results suggest that IR+X-ray AGN may not be the same population
as the IR-only AGN, but we are too limited by our sample to make any
assertion.

To measure the fraction of IR-AGN and test for evolution, we compare
two complete samples of cluster galaxies: 1) an optically-selected
sample with members brighter than $V_{\mathrm{AB}} = -21.5$
(rest-frame) and 2) a mid-IR selected sample with members brighter
than ($M^{\star}(z)[3.6\mu$m$]+0.5$) \citep{Muzzin:08} that is
essentially a stellar mass cut.  For the eight galaxy clusters at
$z<1$, we place a strong upper limit of $\lesssim3$\% on the fraction
of IR-AGN for both cluster samples.  Because IR-AGN tend to be hosted
by late-type galaxies and the morphological mix in clusters evolves
\citep{Dressler:97,Postman:05}, we also consider only late-type
members and find that the fraction with IR-AGN is $\lesssim5$\% for
both samples.  These low IR-AGN fractions are surprising given that
the fraction of (dusty) star formation in these same clusters
increases by about a factor of four at $0<z<1$
\citep{Saintonge:08,Bai:10}.  However, an IR-AGN fraction of
$\sim1$\% is consistent with the low fraction of X-ray AGN in galaxy
clusters \citep[$\leq1$\%][]{Martini:09} and the relative populations
of X-ray vs. IR AGN \citep{Hickox:09}.

In contrast, our single galaxy cluster at $z=1.24$ (RDCS 1252) has a
measurably higher IR-AGN fraction of $\sim15$\% (all galaxy types) and
$\sim70$\% (late-types only).  However, RDCS 1252 may simply be an
unusually active cluster.  Also, the IR color selection starts to
suffer stronger contamination from non-AGN members at $z>1.2$.  

We also compare our \agnfrac\ measurements in dense clusters at $z<1$
to that of the Bo\"otes field from the AGES, which probes sparser
galactic environments over a similar range in redshift
($0.25<z<0.8$). Using optical magnitude cuts similar to
this study, \agnfrac\ is measured to be \twid 2\% in the field
(R. Hickox, private communication). Consequently, we do not see a
statistically significant variation in \agnfrac\ here that would be
caused by local galaxy density. However, the question still remains as to
whether or not IR-AGN have a more profound impact at $z>1$.

We note that while the IR color selection successfully identifies
$\sim90$\% of broad-line AGN, it does miss $\sim60$\% of narrow-line AGN
\citep{Stern:05} and so \agnfrac\ is undoubtedly incomplete,
i.e. underestimated.  
%However, it is also quite likely that a number
%of our IR-AGN (up to 50\%) are actually strongly star forming systems
%\citep{Donley:08}, 
%
% addition begin
On the other hand, contamination may arise from star forming
galaxies falsely identified as IR-AGN; thus we may also be overestimating
\agnfrac. Such contamination can be as high as 20--50\%
\citep{Donley:08,Hickox:09} and is more influential at low
luminosities. For simplicity, we ignore these two competing effects
because they are beyond the scope of our analysis and do not change
our general conclusions.
% addition end

Taken as a whole, our results show that IR-AGN and star formation are
not strongly correlated at $z<1$ because the IR-AGN fraction is
uniformly very low ($\sim1$\%) at $z<1$ whereas several of these
clusters have star forming fractions of $\gtrsim20$\%
\citep{Saintonge:08}.  We do find a hint of evolution in the IR-AGN
fraction at $z\sim1.2$, but only with a more extensive mid-IR survey
of galaxy clusters, particularly at $z>1$, can we confirm this
intriguing result.  A study
%of the field
with equally deep
spectroscopic coverage across this wide redshift range is also needed
to robustly measure \agnfrac\ and test for evolution as a function of
environment.  While \textit{Spitzer}/IRAC is no longer available, such
studies should be possible with the upcoming public survey by the {\it
Wide-field Infrared Survey Explorer} (WISE).
%and targeted observations with the {\it Herschel Space Telescope}.

\acknowledgements

We thank our referee, Ryan Hickox, as well as Paul Martini and David
Atlee for their instructive comments and suggestions that improved our
paper. We would also like to thank Pauline Barmby for assistance with the
IRAC photometry, Julien Devriendt for providing template galactic
SEDs, Ricardo Demarco for providing \textit{HST} imaging of RDCS 1252
and Adam Muzzin for providing data on the 3.6\um\ luminosity
function. K.T. acknowledges generous support from the Swiss National
Science Foundation (grant PP002-110576).  In this work we have made
use of the kcorrect software originally developed by Mike Blanton
\citep{Blanton:07} and adapted to Python by Taro Sato. This work is
based on observations made by the \textit{Spitzer Space Telescope}
which is operated by the Jet Propulsion Laboratory, California
Institute of Technology, under contract with the NASA. This research
has made use of the VizieR catalogue access tool, CDS, Strasbourg,
France.

\nocite{*}
\bibliographystyle{apj}
\bibliography{mybib}

\begin{thebibliography}{}

\bibitem[\protect\citeauthoryear{{Allen} \& {Fabian}}{{Allen} \&
  {Fabian}}{1998}]{Allen:98}
{Allen}, S.~W.,  \& {Fabian}, A.~C. 1998, \mnras, 297, L63

\bibitem[\protect\citeauthoryear{{Andreani} et~al.}{{Andreani}
  et~al.}{2003}]{Andreani:03}
{Andreani}, P., {Cristiani}, S., {Grazian}, A., {La Franca}, F.,  \&
  {Goldschmidt}, P. 2003, \aj, 125, 444

\bibitem[\protect\citeauthoryear{{Arabadjis}, {Bautz}, \&
  {Garmire}}{{Arabadjis} et~al.}{2002}]{Arabadjis:02}
{Arabadjis}, J.~S., {Bautz}, M.~W.,  \& {Garmire}, G.~P. 2002, \apj, 572, 66

\bibitem[\protect\citeauthoryear{{Arnold} et~al.}{{Arnold}
  et~al.}{2009}]{Arnold:09}
{Arnold}, T.~J., {Martini}, P., {Mulchaey}, J.~S., {Berti}, A.,  \& {Jeltema},
  T.~E. 2009, \apj, 707, 1691

\bibitem[\protect\citeauthoryear{Ashby et~al.}{Ashby et~al.}{2009}]{Ashby:09}
Ashby, M., et~al. 2009, Astrophys.J., 701, 428

\bibitem[\protect\citeauthoryear{{Atlee} et~al.}{{Atlee}
  et~al.}{2011}]{Atlee:11}
{Atlee}, D.~W., {Martini}, P., {Assef}, R.~J., {Kelson}, D.~D.,  \& {Mulchaey},
  J.~S. 2011, \apj, 729, 22

\bibitem[\protect\citeauthoryear{{Bai} et~al.}{{Bai} et~al.}{2010}]{Bai:10}
{Bai}, L., {Rasmussen}, J., {Mulchaey}, J.~S., {Dariush}, A., {Raychaudhury},
  S.,  \& {Ponman}, T.~J. 2010, \apj, 713, 637

\bibitem[\protect\citeauthoryear{{Bardeau} et~al.}{{Bardeau}
  et~al.}{2007}]{Bardeau:07}
{Bardeau}, S., {Soucail}, G., {Kneib}, J., {Czoske}, O., {Ebeling}, H.,
  {Hudelot}, P., {Smail}, I.,  \& {Smith}, G.~P. 2007, \aap, 470, 449

\bibitem[\protect\citeauthoryear{{Barmby} et~al.}{{Barmby}
  et~al.}{2006}]{Barmby:06}
{Barmby}, P., et~al. 2006, \apj, 642, 126

\bibitem[\protect\citeauthoryear{{Barmby} et~al.}{{Barmby}
  et~al.}{2008}]{Barmby:08}
{Barmby}, P., {Huang}, J., {Ashby}, M.~L.~N., {Eisenhardt}, P.~R.~M., {Fazio},
  G.~G., {Willner}, S.~P.,  \& {Wright}, E.~L. 2008, \apjs, 177, 431

\bibitem[\protect\citeauthoryear{{Bertin} \& {Arnouts}}{{Bertin} \&
  {Arnouts}}{1996}]{Bertin:96}
{Bertin}, E.,  \& {Arnouts}, S. 1996, \aaps, 117, 393

\bibitem[\protect\citeauthoryear{{Best} et~al.}{{Best} et~al.}{2002}]{Best:02}
{Best}, P.~N., {van Dokkum}, P.~G., {Franx}, M.,  \& {R{\"o}ttgering}, H.~J.~A.
  2002, \mnras, 330, 17

\bibitem[\protect\citeauthoryear{{Blakeslee} et~al.}{{Blakeslee}
  et~al.}{2003}]{Blakeslee:03}
{Blakeslee}, J.~P., et~al. 2003, \apjl, 596, L143

\bibitem[\protect\citeauthoryear{{Blakeslee} et~al.}{{Blakeslee}
  et~al.}{2006}]{Blakeslee:06}
{Blakeslee}, J.~P., et~al. 2006, \apj, 644, 30

\bibitem[\protect\citeauthoryear{{Blanton} \& {Roweis}}{{Blanton} \&
  {Roweis}}{2007}]{Blanton:07}
{Blanton}, M.~R.,  \& {Roweis}, S. 2007, \aj, 133, 734

\bibitem[\protect\citeauthoryear{{Bower} et~al.}{{Bower}
  et~al.}{2006}]{Bower:06}
{Bower}, R.~G., {Benson}, A.~J., {Malbon}, R., {Helly}, J.~C., {Frenk}, C.~S.,
  {Baugh}, C.~M., {Cole}, S.,  \& {Lacey}, C.~G. 2006, \mnras, 370, 645

\bibitem[\protect\citeauthoryear{{Bower}, {Lucey}, \& {Ellis}}{{Bower}
  et~al.}{1992}]{Bower:92}
{Bower}, R.~G., {Lucey}, J.~R.,  \& {Ellis}, R.~S. 1992, \mnras, 254, 589

\bibitem[\protect\citeauthoryear{{Bower}, {McCarthy}, \& {Benson}}{{Bower}
  et~al.}{2008}]{Bower:08}
{Bower}, R.~G., {McCarthy}, I.~G.,  \& {Benson}, A.~J. 2008, \mnras, 390, 1399

\bibitem[\protect\citeauthoryear{{Capak} et~al.}{{Capak}
  et~al.}{2007}]{Capak:07}
{Capak}, P., {Abraham}, R.~G., {Ellis}, R.~S., {Mobasher}, B., {Scoville}, N.,
  {Sheth}, K.,  \& {Koekemoer}, A. 2007, \apjs, 172, 284

\bibitem[\protect\citeauthoryear{{Chung} et~al.}{{Chung}
  et~al.}{2010}]{Sun_Chung:10}
{Chung}, S.~M., {Gonzalez}, A.~H., {Clowe}, D., {Markevitch}, M.,  \&
  {Zaritsky}, D. 2010, \apj, 725, 1536

\bibitem[\protect\citeauthoryear{{Clowe} et~al.}{{Clowe}
  et~al.}{2006}]{Clowe:06}
{Clowe}, D., {Brada{\v c}}, M., {Gonzalez}, A.~H., {Markevitch}, M., {Randall},
  S.~W., {Jones}, C.,  \& {Zaritsky}, D. 2006, \apjl, 648, L109

\bibitem[\protect\citeauthoryear{{Cooray} et~al.}{{Cooray}
  et~al.}{1998}]{Cooray:98}
{Cooray}, A.~R., {Grego}, L., {Holzapfel}, W.~L., {Joy}, M.,  \& {Carlstrom},
  J.~E. 1998, \aj, 115, 1388

\bibitem[\protect\citeauthoryear{{Croton} et~al.}{{Croton}
  et~al.}{2006}]{Croton:06}
{Croton}, D.~J., et~al. 2006, \mnras, 365, 11

\bibitem[\protect\citeauthoryear{{Czoske} et~al.}{{Czoske}
  et~al.}{2001}]{Czoske:01}
{Czoske}, O., {Kneib}, J., {Soucail}, G., {Bridges}, T.~J., {Mellier}, Y.,  \&
  {Cuillandre}, J. 2001, \aap, 372, 391

\bibitem[\protect\citeauthoryear{{Czoske} et~al.}{{Czoske}
  et~al.}{2002}]{Czoske:02}
{Czoske}, O., {Moore}, B., {Kneib}, J.,  \& {Soucail}, G. 2002, \aap, 386, 31

\bibitem[\protect\citeauthoryear{{De Lucia} et~al.}{{De Lucia}
  et~al.}{2007}]{DeLucia:07}
{De Lucia}, G., et~al. 2007, \mnras, 374, 809

\bibitem[\protect\citeauthoryear{{de Propris} et~al.}{{de Propris}
  et~al.}{1999}]{dePropris:99}
{de Propris}, R., {Stanford}, S.~A., {Eisenhardt}, P.~R., {Dickinson}, M.,  \&
  {Elston}, R. 1999, \aj, 118, 719

\bibitem[\protect\citeauthoryear{{Demarco} et~al.}{{Demarco}
  et~al.}{2007}]{Demarco:07}
{Demarco}, R., et~al. 2007, \apj, 663, 164

\bibitem[\protect\citeauthoryear{{Demarco} et~al.}{{Demarco}
  et~al.}{2005}]{Demarco:05}
{Demarco}, R., et~al. 2005, \aap, 432, 381

\bibitem[\protect\citeauthoryear{{Devriendt}, {Guiderdoni}, \&
  {Sadat}}{{Devriendt} et~al.}{1999}]{Devriendt:99}
{Devriendt}, J.~E.~G., {Guiderdoni}, B.,  \& {Sadat}, R. 1999, \aap, 350, 381

\bibitem[\protect\citeauthoryear{{Donahue} et~al.}{{Donahue}
  et~al.}{2003}]{Donahue:03}
{Donahue}, M., {Gaskin}, J.~A., {Patel}, S.~K., {Joy}, M., {Clowe}, D.,  \&
  {Hughes}, J.~P. 2003, \apj, 598, 190

\bibitem[\protect\citeauthoryear{{Donahue} et~al.}{{Donahue}
  et~al.}{1999}]{Donahue:99}
{Donahue}, M., {Voit}, G.~M., {Scharf}, C.~A., {Gioia}, I.~M., {Mullis}, C.~R.,
  {Hughes}, J.~P.,  \& {Stocke}, J.~T. 1999, \apj, 527, 525

\bibitem[\protect\citeauthoryear{{Donley} et~al.}{{Donley}
  et~al.}{2008}]{Donley:08}
{Donley}, J.~L., {Rieke}, G.~H., {P{\'e}rez-Gonz{\'a}lez}, P.~G.,  \& {Barro},
  G. 2008, \apj, 687, 111

\bibitem[\protect\citeauthoryear{{Dressler} et~al.}{{Dressler}
  et~al.}{1997}]{Dressler:97}
{Dressler}, A., et~al. 1997, \apj, 490, 577

\bibitem[\protect\citeauthoryear{{Dressler} \& {Shectman}}{{Dressler} \&
  {Shectman}}{1988}]{Dressler:88}
{Dressler}, A.,  \& {Shectman}, S.~A. 1988, \aj, 95, 985

\bibitem[\protect\citeauthoryear{{Duc} et~al.}{{Duc} et~al.}{2002}]{Duc:02}
{Duc}, P., et~al. 2002, \aap, 382, 60

\bibitem[\protect\citeauthoryear{{Dye} et~al.}{{Dye} et~al.}{2001}]{Dye:01}
{Dye}, S., {Taylor}, A.~N., {Thommes}, E.~M., {Meisenheimer}, K., {Wolf}, C.,
  \& {Peacock}, J.~A. 2001, \mnras, 321, 685

\bibitem[\protect\citeauthoryear{{Eastman} et~al.}{{Eastman}
  et~al.}{2007}]{Eastman:07}
{Eastman}, J., {Martini}, P., {Sivakoff}, G., {Kelson}, D.~D., {Mulchaey},
  J.~S.,  \& {Tran}, K. 2007, \apjl, 664, L9

\bibitem[\protect\citeauthoryear{{Eckart} et~al.}{{Eckart}
  et~al.}{2010}]{Eckart:10}
{Eckart}, M.~E., {McGreer}, I.~D., {Stern}, D., {Harrison}, F.~A.,  \&
  {Helfand}, D.~J. 2010, \apj, 708, 584

\bibitem[\protect\citeauthoryear{{Eisenhardt} et~al.}{{Eisenhardt}
  et~al.}{2004}]{Eisenhardt:04}
{Eisenhardt}, P.~R., et~al. 2004, \apjs, 154, 48

\bibitem[\protect\citeauthoryear{{Elvis} et~al.}{{Elvis}
  et~al.}{1994}]{Elvis:94}
{Elvis}, M., et~al. 1994, \apjs, 95, 1

\bibitem[\protect\citeauthoryear{{Fabricant}, {Franx}, \& {van
  Dokkum}}{{Fabricant} et~al.}{2000}]{Fabricant:00}
{Fabricant}, D., {Franx}, M.,  \& {van Dokkum}, P. 2000, \apj, 539, 577

\bibitem[\protect\citeauthoryear{{Farouki} \& {Shapiro}}{{Farouki} \&
  {Shapiro}}{1981}]{Farouki:81}
{Farouki}, R.,  \& {Shapiro}, S.~L. 1981, \apj, 243, 32

\bibitem[\protect\citeauthoryear{{Fasano} et~al.}{{Fasano}
  et~al.}{2000}]{Fasano:00}
{Fasano}, G., {Poggianti}, B.~M., {Couch}, W.~J., {Bettoni}, D.,
  {Kj{\ae}rgaard}, P.,  \& {Moles}, M. 2000, \apj, 542, 673

\bibitem[\protect\citeauthoryear{{Fazio} et~al.}{{Fazio}
  et~al.}{2004}]{Fazio:04}
{Fazio}, G.~G., et~al. 2004, \apjs, 154, 10

\bibitem[\protect\citeauthoryear{{Fisher} et~al.}{{Fisher}
  et~al.}{1998}]{Fisher:98}
{Fisher}, D., {Fabricant}, D., {Franx}, M.,  \& {van Dokkum}, P. 1998, \apj,
  498, 195

\bibitem[\protect\citeauthoryear{{Fontanot} et~al.}{{Fontanot}
  et~al.}{2010}]{Fontanot:10}
{Fontanot}, F., {Pasquali}, A., {De Lucia}, G., {van den Bosch}, F.~C.,
  {Somerville}, R.~S.,  \& {Kang}, X. 2010, ArXiv e-prints

\bibitem[\protect\citeauthoryear{{Gabor} et~al.}{{Gabor}
  et~al.}{2010}]{Gabor:10}
{Gabor}, J.~M., {Dav{\'e}}, R., {Finlator}, K.,  \& {Oppenheimer}, B.~D. 2010,
  \mnras, 407, 749

\bibitem[\protect\citeauthoryear{{Galametz} et~al.}{{Galametz}
  et~al.}{2009}]{Galametz:09}
{Galametz}, A., et~al. 2009, \apj, 694, 1309

\bibitem[\protect\citeauthoryear{{Gandhi} et~al.}{{Gandhi}
  et~al.}{2009}]{Gandhi:09}
{Gandhi}, P., {Horst}, H., {Smette}, A., {H{\"o}nig}, S., {Comastri}, A.,
  {Gilli}, R., {Vignali}, C.,  \& {Duschl}, W. 2009, \aap, 502, 457

\bibitem[\protect\citeauthoryear{{Gehrels}}{{Gehrels}}{1986}]{Gehrels:86}
{Gehrels}, N. 1986, \apj, 303, 336

\bibitem[\protect\citeauthoryear{{Gilmour} et~al.}{{Gilmour}
  et~al.}{2007}]{Gilmour:07}
{Gilmour}, R., {Gray}, M.~E., {Almaini}, O., {Best}, P., {Wolf}, C.,
  {Meisenheimer}, K., {Papovich}, C.,  \& {Bell}, E. 2007, \mnras, 380, 1467

\bibitem[\protect\citeauthoryear{{Griffith} \& {Stern}}{{Griffith} \&
  {Stern}}{2010}]{Griffith:10}
{Griffith}, R.~L.,  \& {Stern}, D. 2010, \aj, 140, 533

\bibitem[\protect\citeauthoryear{{Gunn} \& {Gott}}{{Gunn} \&
  {Gott}}{1972}]{Gunn:72}
{Gunn}, J.~E.,  \& {Gott}, J.~R., III. 1972, \apj, 176, 1

\bibitem[\protect\citeauthoryear{{Hart}, {Stocke}, \& {Hallman}}{{Hart}
  et~al.}{2009}]{Hart:09}
{Hart}, Q.~N., {Stocke}, J.~T.,  \& {Hallman}, E.~J. 2009, \apj, 705, 854

\bibitem[\protect\citeauthoryear{{Hickox} et~al.}{{Hickox}
  et~al.}{2007}]{Hickox:07}
{Hickox}, R.~C., et~al. 2007, \apj, 671, 1365

\bibitem[\protect\citeauthoryear{{Hickox} et~al.}{{Hickox}
  et~al.}{2009}]{Hickox:09}
{Hickox}, R.~C., et~al. 2009, \apj, 696, 891

\bibitem[\protect\citeauthoryear{{Hilton} et~al.}{{Hilton}
  et~al.}{2010}]{Hilton:10}
{Hilton}, M., et~al. 2010, \apj, 718, 133

\bibitem[\protect\citeauthoryear{{Hoekstra} et~al.}{{Hoekstra}
  et~al.}{1998}]{Hoekstra:98}
{Hoekstra}, H., {Franx}, M., {Kuijken}, K.,  \& {Squires}, G. 1998, \apj, 504,
  636

\bibitem[\protect\citeauthoryear{{Hogg} et~al.}{{Hogg} et~al.}{2004}]{Hogg:04}
{Hogg}, D.~W., et~al. 2004, \apjl, 601, L29

\bibitem[\protect\citeauthoryear{{Holden} et~al.}{{Holden}
  et~al.}{2007}]{Holden:07}
{Holden}, B.~P., et~al. 2007, \apj, 670, 190

\bibitem[\protect\citeauthoryear{{Hopkins} \& {Hernquist}}{{Hopkins} \&
  {Hernquist}}{2006}]{Hopkins:06}
{Hopkins}, P.~F.,  \& {Hernquist}, L. 2006, \apjs, 166, 1

\bibitem[\protect\citeauthoryear{{Hopkins} et~al.}{{Hopkins}
  et~al.}{2008}]{Hopkins:08}
{Hopkins}, P.~F., {Hernquist}, L., {Cox}, T.~J.,  \& {Kere{\v s}}, D. 2008,
  \apjs, 175, 356

\bibitem[\protect\citeauthoryear{{Hopkins} et~al.}{{Hopkins}
  et~al.}{2009}]{Hopkins:09}
{Hopkins}, P.~F., {Hickox}, R., {Quataert}, E.,  \& {Hernquist}, L. 2009,
  \mnras, 398, 333

\bibitem[\protect\citeauthoryear{{Jee} et~al.}{{Jee} et~al.}{2005a}]{Jee:05a}
{Jee}, M.~J., {White}, R.~L., {Ben{\'{\i}}tez}, N., {Ford}, H.~C., {Blakeslee},
  J.~P., {Rosati}, P., {Demarco}, R.,  \& {Illingworth}, G.~D. 2005a, \apj,
  618, 46

\bibitem[\protect\citeauthoryear{{Jee} et~al.}{{Jee} et~al.}{2005b}]{Jee:05b}
{Jee}, M.~J., {White}, R.~L., {Ford}, H.~C., {Blakeslee}, J.~P., {Illingworth},
  G.~D., {Coe}, D.~A.,  \& {Tran}, K. 2005b, \apj, 634, 813

\bibitem[\protect\citeauthoryear{{Johnson}, {Best}, \& {Almaini}}{{Johnson}
  et~al.}{2003}]{Johnson:03}
{Johnson}, O., {Best}, P.~N.,  \& {Almaini}, O. 2003, \mnras, 343, 924

\bibitem[\protect\citeauthoryear{{Kneib} et~al.}{{Kneib}
  et~al.}{2003}]{Kneib:03}
{Kneib}, J., et~al. 2003, \apj, 598, 804

\bibitem[\protect\citeauthoryear{{Kocevski} et~al.}{{Kocevski}
  et~al.}{2009}]{Kocevski:09}
{Kocevski}, D.~D., {Lubin}, L.~M., {Lemaux}, B.~C., {Gal}, R.~R., {Fassnacht},
  C.~D.,  \& {Squires}, G.~K. 2009, \apjl, 703, L33

\bibitem[\protect\citeauthoryear{{Kuraszkiewicz} et~al.}{{Kuraszkiewicz}
  et~al.}{2003}]{Kuraszkiewicz:03}
{Kuraszkiewicz}, J.~K., et~al. 2003, \apj, 590, 128

\bibitem[\protect\citeauthoryear{Lacy et~al.}{Lacy et~al.}{2007}]{Lacy:07}
Lacy, M., Petric, A., Sajina, A., Canalizo, G., Storrie-Lombardi, L., Armus,
  L., Fadda, D.,  \& Marleau, F. 2007, Astron.J., 133, 186

\bibitem[\protect\citeauthoryear{{Lacy} et~al.}{{Lacy} et~al.}{2004}]{Lacy:04}
{Lacy}, M., et~al. 2004, \apjs, 154, 166

\bibitem[\protect\citeauthoryear{{Lacy} et~al.}{{Lacy} et~al.}{2005}]{Lacy:05}
{Lacy}, M., et~al. 2005, \apjs, 161, 41

\bibitem[\protect\citeauthoryear{{Lagos}, {Cora}, \& {Padilla}}{{Lagos}
  et~al.}{2008}]{Lagos:08}
{Lagos}, C.~D.~P., {Cora}, S.~A.,  \& {Padilla}, N.~D. 2008, \mnras, 388, 587

\bibitem[\protect\citeauthoryear{{Lombardi} et~al.}{{Lombardi}
  et~al.}{2005}]{Lombardi:05}
{Lombardi}, M., et~al. 2005, \apj, 623, 42

\bibitem[\protect\citeauthoryear{{Lubin} et~al.}{{Lubin}
  et~al.}{2009}]{Lubin:09}
{Lubin}, L.~M., {Gal}, R.~R., {Lemaux}, B.~C., {Kocevski}, D.~D.,  \&
  {Squires}, G.~K. 2009, \aj, 137, 4867

\bibitem[\protect\citeauthoryear{{Lutz} et~al.}{{Lutz} et~al.}{2004}]{Lutz:04}
{Lutz}, D., {Maiolino}, R., {Spoon}, H.~W.~W.,  \& {Moorwood}, A.~F.~M. 2004,
  \aap, 418, 465

\bibitem[\protect\citeauthoryear{{Mahajan}, {Haines}, \&
  {Raychaudhury}}{{Mahajan} et~al.}{2010}]{Mahajan:10}
{Mahajan}, S., {Haines}, C.~P.,  \& {Raychaudhury}, S. 2010, \mnras, 404, 1745

\bibitem[\protect\citeauthoryear{{Makovoz} et~al.}{{Makovoz}
  et~al.}{2006}]{Makovoz:06}
{Makovoz}, D., {Roby}, T., {Khan}, I.,  \& {Booth}, H. 2006, in Presented at
  the Society of Photo-Optical Instrumentation Engineers (SPIE) Conference,
  Vol. 6274, Society of Photo-Optical Instrumentation Engineers (SPIE)
  Conference Series

\bibitem[\protect\citeauthoryear{{Martel} et~al.}{{Martel}
  et~al.}{2007}]{Martel:07}
{Martel}, A.~R., {Menanteau}, F., {Tozzi}, P., {Ford}, H.~C.,  \& {Infante}, L.
  2007, \apjs, 168, 19

\bibitem[\protect\citeauthoryear{{Martini}, {Mulchaey}, \& {Kelson}}{{Martini}
  et~al.}{2007}]{Martini:07}
{Martini}, P., {Mulchaey}, J.~S.,  \& {Kelson}, D.~D. 2007, \apj, 664, 761

\bibitem[\protect\citeauthoryear{Martini, Sivakoff, \& Mulchaey}{Martini
  et~al.}{2009}]{Martini:09}
Martini, P., Sivakoff, G.~R.,  \& Mulchaey, J.~S. 2009, Astrophys.J., 701, 66

\bibitem[\protect\citeauthoryear{{McCarthy} et~al.}{{McCarthy}
  et~al.}{2010}]{McCarthy:10}
{McCarthy}, I.~G., et~al. 2010, \mnras, 406, 822

\bibitem[\protect\citeauthoryear{{Michard} \& {Andreon}}{{Michard} \&
  {Andreon}}{2008}]{Michard:08}
{Michard}, R.,  \& {Andreon}, S. 2008, \aap, 490, 923

\bibitem[\protect\citeauthoryear{{Mobasher} et~al.}{{Mobasher}
  et~al.}{2001}]{Mobasher:01}
{Mobasher}, B., et~al. 2001, \apjs, 137, 279

\bibitem[\protect\citeauthoryear{{Molnar} et~al.}{{Molnar}
  et~al.}{2002}]{Molnar:02}
{Molnar}, S.~M., {Hughes}, J.~P., {Donahue}, M.,  \& {Joy}, M. 2002, \apjl,
  573, L91

\bibitem[\protect\citeauthoryear{{Moran} et~al.}{{Moran}
  et~al.}{2005}]{Moran:05}
{Moran}, S.~M., {Ellis}, R.~S., {Treu}, T., {Smail}, I., {Dressler}, A.,
  {Coil}, A.~L.,  \& {Smith}, G.~P. 2005, \apj, 634, 977

\bibitem[\protect\citeauthoryear{{Moran} et~al.}{{Moran}
  et~al.}{2007a}]{Moran1:07}
{Moran}, S.~M., {Ellis}, R.~S., {Treu}, T., {Smith}, G.~P., {Rich}, R.~M.,  \&
  {Smail}, I. 2007a, \apj, 671, 1503

\bibitem[\protect\citeauthoryear{{Moran} et~al.}{{Moran}
  et~al.}{2007b}]{Moran2:07}
{Moran}, S.~M., {Miller}, N., {Treu}, T., {Ellis}, R.~S.,  \& {Smith}, G.~P.
  2007b, \apj, 659, 1138

\bibitem[\protect\citeauthoryear{{Muzzin} et~al.}{{Muzzin}
  et~al.}{2008}]{Muzzin:08}
{Muzzin}, A., {Wilson}, G., {Lacy}, M., {Yee}, H.~K.~C.,  \& {Stanford}, S.~A.
  2008, \apj, 686, 966

\bibitem[\protect\citeauthoryear{{Papovich} et~al.}{{Papovich}
  et~al.}{2006}]{Papovich:06}
{Papovich}, C., et~al. 2006, \aj, 132, 231

\bibitem[\protect\citeauthoryear{{Park} et~al.}{{Park} et~al.}{2010}]{Park:10}
{Park}, S.~Q., et~al. 2010, \apj, 717, 1181

\bibitem[\protect\citeauthoryear{{Pier} \& {Krolik}}{{Pier} \&
  {Krolik}}{1992}]{Pier:92}
{Pier}, E.~A.,  \& {Krolik}, J.~H. 1992, \apj, 401, 99

\bibitem[\protect\citeauthoryear{{Polletta} et~al.}{{Polletta}
  et~al.}{2007}]{Polletta:07}
{Polletta}, M., et~al. 2007, \apj, 663, 81

\bibitem[\protect\citeauthoryear{{Postman} et~al.}{{Postman}
  et~al.}{2005}]{Postman:05}
{Postman}, M., et~al. 2005, \apj, 623, 721

\bibitem[\protect\citeauthoryear{{Puchwein}, {Sijacki}, \&
  {Springel}}{{Puchwein} et~al.}{2008}]{Puchwein:08}
{Puchwein}, E., {Sijacki}, D.,  \& {Springel}, V. 2008, \apjl, 687, L53

\bibitem[\protect\citeauthoryear{Reach et~al.}{Reach et~al.}{2005}]{Reach:05}
Reach, W.~T., et~al. 2005, Publ.Astron.Soc.Pac., 117, 978

\bibitem[\protect\citeauthoryear{{Richstone}}{{Richstone}}{1976}]{Richstone:76}
{Richstone}, D.~O. 1976, \apj, 204, 642

\bibitem[\protect\citeauthoryear{{Rines} et~al.}{{Rines}
  et~al.}{2003}]{Rines:03}
{Rines}, K., {Geller}, M.~J., {Kurtz}, M.~J.,  \& {Diaferio}, A. 2003, \aj,
  126, 2152

\bibitem[\protect\citeauthoryear{{Rosati} et~al.}{{Rosati}
  et~al.}{2004}]{Rosati:04}
{Rosati}, P., et~al. 2004, \aj, 127, 230

\bibitem[\protect\citeauthoryear{{Rudnick} et~al.}{{Rudnick}
  et~al.}{2009}]{Rudnick:09}
{Rudnick}, G., et~al. 2009, \apj, 700, 1559

\bibitem[\protect\citeauthoryear{{Sacchi} et~al.}{{Sacchi}
  et~al.}{2009}]{Sacchi:09}
{Sacchi}, N., et~al. 2009, \apj, 703, 1778

\bibitem[\protect\citeauthoryear{{Saintonge}, {Tran}, \& {Holden}}{{Saintonge}
  et~al.}{2008}]{Saintonge:08}
{Saintonge}, A., {Tran}, K.,  \& {Holden}, B.~P. 2008, \apjl, 685, L113

\bibitem[\protect\citeauthoryear{{Sandage} \& {Visvanathan}}{{Sandage} \&
  {Visvanathan}}{1978}]{Sandage:78}
{Sandage}, A.,  \& {Visvanathan}, N. 1978, \apj, 223, 707

\bibitem[\protect\citeauthoryear{{Sanders} et~al.}{{Sanders}
  et~al.}{1989}]{Sanders:89}
{Sanders}, D.~B., {Phinney}, E.~S., {Neugebauer}, G., {Soifer}, B.~T.,  \&
  {Matthews}, K. 1989, \apj, 347, 29

\bibitem[\protect\citeauthoryear{{Schechter}}{{Schechter}}{1976}]{Schechter:76}
{Schechter}, P. 1976, \apj, 203, 297

\bibitem[\protect\citeauthoryear{{Soifer} \& {First Look Survey Team}}{{Soifer}
  \& {First Look Survey Team}}{2004}]{Soifer:04}
{Soifer}, B.~T.,  \& {First Look Survey Team}. 2004, in Bulletin of the
  American Astronomical Society, Vol.~36, Bulletin of the American Astronomical
  Society, 699

\bibitem[\protect\citeauthoryear{{Stern} et~al.}{{Stern}
  et~al.}{2005}]{Stern:05}
{Stern}, D., et~al. 2005, \apj, 631, 163

\bibitem[\protect\citeauthoryear{{Stott} et~al.}{{Stott}
  et~al.}{2007}]{Stott:07}
{Stott}, J.~P., {Smail}, I., {Edge}, A.~C., {Ebeling}, H., {Smith}, G.~P.,
  {Kneib}, J.,  \& {Pimbblet}, K.~A. 2007, \apj, 661, 95

\bibitem[\protect\citeauthoryear{{Teyssier} et~al.}{{Teyssier}
  et~al.}{2010}]{Teyssier:10}
{Teyssier}, R., {Moore}, B., {Martizzi}, D., {Dubois}, Y.,  \& {Mayer}, L.
  2010, ArXiv e-prints

\bibitem[\protect\citeauthoryear{{Thompson} et~al.}{{Thompson}
  et~al.}{2009}]{Thompson:09}
{Thompson}, G.~D., {Levenson}, N.~A., {Uddin}, S.~A.,  \& {Sirocky}, M.~M.
  2009, \apj, 697, 182

\bibitem[\protect\citeauthoryear{{Tran} et~al.}{{Tran} et~al.}{2007}]{Tran:07}
{Tran}, K., {Franx}, M., {Illingworth}, G.~D., {van Dokkum}, P., {Kelson},
  D.~D., {Blakeslee}, J.~P.,  \& {Postman}, M. 2007, \apj, 661, 750

\bibitem[\protect\citeauthoryear{{Tran} et~al.}{{Tran} et~al.}{2010}]{Tran:10}
{Tran}, K., et~al. 2010, \apjl, 719, L126

\bibitem[\protect\citeauthoryear{{Tran} et~al.}{{Tran}
  et~al.}{2005a}]{Tran:05a}
{Tran}, K., {van Dokkum}, P., {Franx}, M., {Illingworth}, G.~D., {Kelson},
  D.~D.,  \& {Schreiber}, N.~M.~F. 2005a, \apjl, 627, L25

\bibitem[\protect\citeauthoryear{{Tran} et~al.}{{Tran}
  et~al.}{2005b}]{Tran:05b}
{Tran}, K., {van Dokkum}, P., {Illingworth}, G.~D., {Kelson}, D., {Gonzalez},
  A.,  \& {Franx}, M. 2005b, \apj, 619, 134

\bibitem[\protect\citeauthoryear{{Treu} et~al.}{{Treu} et~al.}{2003}]{Treu:03}
{Treu}, T., {Ellis}, R.~S., {Kneib}, J., {Dressler}, A., {Smail}, I., {Czoske},
  O., {Oemler}, A.,  \& {Natarajan}, P. 2003, \apj, 591, 53

\bibitem[\protect\citeauthoryear{{van Dokkum} et~al.}{{van Dokkum}
  et~al.}{1998}]{vanDokkum:98}
{van Dokkum}, P.~G., {Franx}, M., {Kelson}, D.~D., {Illingworth}, G.~D.,
  {Fisher}, D.,  \& {Fabricant}, D. 1998, \apj, 500, 714

\bibitem[\protect\citeauthoryear{{Verdugo}, {de Diego}, \&
  {Limousin}}{{Verdugo} et~al.}{2007}]{Verdugo:07}
{Verdugo}, T., {de Diego}, J.~A.,  \& {Limousin}, M. 2007, \apj, 664, 702

\bibitem[\protect\citeauthoryear{{Werner} et~al.}{{Werner}
  et~al.}{2004}]{Werner:04}
{Werner}, M.~W., et~al. 2004, \apjs, 154, 1

\bibitem[\protect\citeauthoryear{{Zhang} et~al.}{{Zhang}
  et~al.}{2004}]{Zhang:04}
{Zhang}, F., {Han}, Z., {Li}, L.,  \& {Hurley}, J.~R. 2004, \mnras, 350, 710

\bibitem[\protect\citeauthoryear{{Zhang} et~al.}{{Zhang}
  et~al.}{2005}]{Zhang:05}
{Zhang}, Y., {B{\"o}hringer}, H., {Mellier}, Y., {Soucail}, G.,  \& {Forman},
  W. 2005, \aap, 429, 85

\end{thebibliography}

\end{document}